\documentclass[useAMS,usenatbib]{mn2e}
\usepackage{graphicx,natbib,times}

 \title[Evidence for AGN feedback in early-type galaxies]{Observational evidence for AGN feedback in early-type galaxies} \author[Kevin Schawinski et al.]{
  \parbox[t]{16cm}{Kevin~Schawinski$^{1}$\thanks{E-mail: kevins@astro.ox.ac.uk (KS)},
  Daniel Thomas$^{2,1}$,
  Marc Sarzi$^{3,1}$,
  Claudia Maraston$^{2,1}$,
  Sugata Kaviraj$^{1}$,
  Seok-Joo Joo$^{4}$, Sukyoung K. Yi$^{4}$\vspace*{6pt} and Joseph Silk$^{1}$}\\
  $^{1}$Department of Physics, University of Oxford, Oxford OX1 3RH, UK\\
  $^{2}$Institute of Cosmology \& Gravitation, University of Portsmouth, Portsmouth, PO1 2EG, UK\\
  $^{3}$Centre for Astrophysics Research, Science \& Technology
  Research Institute, University of Hertfordshire, Hatfield, UK\\
  $^{4}$Center for Space Astrophysics, Yonsei University, Seoul 120-749,
  Korea\\
}

\begin{document}

\newcommand\aj{{AJ}}%
\newcommand\actaa{{Acta Astron.}}%
\newcommand\araa{{ARA\&A}}%
\newcommand\apj{{ApJ}}%
\newcommand\apjl{{ApJ}}%
\newcommand\apjs{{ApJS}}%
\newcommand\ao{{Appl.~Opt.}}%
\newcommand\apss{{Ap\&SS}}%
\newcommand\aap{{A\&A}}%
\newcommand\aapr{{A\&A~Rev.}}%
\newcommand\aaps{{A\&AS}}%
\newcommand\azh{{AZh}}%
\newcommand\baas{{BAAS}}%
\newcommand\caa{{Chinese Astron. Astrophys.}}%
\newcommand\cjaa{{Chinese J. Astron. Astrophys.}}%
\newcommand\icarus{{Icarus}}%
\newcommand\jcap{{J. Cosmology Astropart. Phys.}}%
\newcommand\jrasc{{JRASC}}%
\newcommand\memras{{MmRAS}}%
\newcommand\mnras{{MNRAS}}%
\newcommand\na{{New A}}%
\newcommand\nar{{New A Rev.}}%
\newcommand\pra{{Phys.~Rev.~A}}%
\newcommand\prb{{Phys.~Rev.~B}}%
\newcommand\prc{{Phys.~Rev.~C}}%
\newcommand\prd{{Phys.~Rev.~D}}%
\newcommand\pre{{Phys.~Rev.~E}}%
\newcommand\prl{{Phys.~Rev.~Lett.}}%
\newcommand\pasa{{PASA}}%
\newcommand\pasp{{PASP}}%
\newcommand\pasj{{PASJ}}%
\newcommand\qjras{{QJRAS}}%
\newcommand\rmxaa{{Rev. Mexicana Astron. Astrofis.}}%
\newcommand\skytel{{S\&T}}%
\newcommand\solphys{{Sol.~Phys.}}%
\newcommand\sovast{{Soviet~Ast.}}%
\newcommand\ssr{{Space~Sci.~Rev.}}%
\newcommand\zap{{ZAp}}%
\newcommand\nat{{Nature}}%
\newcommand\iaucirc{{IAU~Circ.}}%
\newcommand\aplett{{Astrophys.~Lett.}}%
\newcommand\apspr{{Astrophys.~Space~Phys.~Res.}}%
\newcommand\bain{{Bull.~Astron.~Inst.~Netherlands}}%
\newcommand\fcp{{Fund.~Cosmic~Phys.}}%
\newcommand\gca{{Geochim.~Cosmochim.~Acta}}%
\newcommand\grl{{Geophys.~Res.~Lett.}}%
\newcommand\jcp{{J.~Chem.~Phys.}}%
\newcommand\jgr{{J.~Geophys.~Res.}}%
\newcommand\jqsrt{{J.~Quant.~Spec.~Radiat.~Transf.}}%
\newcommand\memsai{{Mem.~Soc.~Astron.~Italiana}}%
\newcommand\nphysa{{Nucl.~Phys.~A}}%
\newcommand\physrep{{Phys.~Rep.}}%
\newcommand\physscr{{Phys.~Scr}}%
\newcommand\planss{{Planet.~Space~Sci.}}%
\newcommand\procspie{{Proc.~SPIE}}%
\newcommand\helvet{{Helvetica~Phys.~Acta}}%

\date{Submitted 2007 August 1, Accepted 2007 September 11.}

\pagerange{\pageref{firstpage}--\pageref{lastpage}} \pubyear{2007}

\maketitle

\label{firstpage}

\begin{abstract}
A major amendment in recent models of hierarchical galaxy formation is
the inclusion of so-called AGN feedback. The energy input from an
active central massive black hole is invoked to suppress star
formation in early-type galaxies at later epochs. A major problem is
that this process is poorly understood, and compelling observational
evidence for its mere existence is still missing. In search for
signatures of AGN feedback, we have compiled a sample of 16,000
early-type galaxies in the redshift range $0.05<z<0.1$ from the SDSS
database (MOSES, Morphologically Selected Ellipticals in SDSS). Key in
our approach is the use of a purely morphological selection criterion
through visual inspection which produces a sample that is not biased
against recent star formation and nuclear activity. Based on the
nebular emission line characteristics we separate between star
formation activity, black hole activity, the composite of the two, and
quiescence. We find that emission is mostly LINER-like in high-mass
galaxies ($\sigma>200\;$km/s) and roughly evenly distributed between
star formation and AGN at intermediate and low ($\sigma<100\;$km/s)
masses. The objects with emission ($\sim 20\;$per cent) are offset
from the red sequence and form a well-defined pattern in the
colour-mass diagram. Star forming early-types inhabit the blue cloud,
while early-types with AGN are located considerably closer to and
almost on the red sequence. Star formation-AGN composites are found
right between these two extremes. We further derive galaxy star
formation histories using a novel method that combines multiwavelength
photometry from near-UV to near-IR and stellar absorption indices.
We find that in those objects deviating from the red sequence star
formation occurred several $100\;$Myr in the past involving
$1-10\;$per cent of the total stellar mass. We identify an
evolutionary sequence from star formation via nuclear activity to
quiescence. This transition process lasts about $1\;$Gyr, and the peak
AGN phase occurs roughly half a Gyr after the starburst. The most
likely interpretation is that star formation is suppressed by nuclear
activity in these objects before they settle on the red sequence. This
is empirical evidence for the occurrence of AGN feedback in early-type
galaxies at recent epochs.
\end{abstract}

\begin{keywords}
galaxies: active, galaxies: elliptical and lenticular,
galaxies: evolution, galaxies: formation, galaxies: active
\end{keywords}

\section{Introduction}

The formation of early-type galaxies is one of the major unsolved
problems of astrophysics. Despite their apparent simplicity, they have
time and again refused to fit into the bigger picture and thus, they
have been the most difficult and illuminating test for galaxy and
structure formation theory. In every generation of theoretical models,
early-type galaxies have helped provide the insights into what physics
was still missing from them and they continue to do so.  The vast
majority of early-type galaxies appear to have formed most of their
stars at high redshift on very short timescales as suggested by the
stellar population properties of nearby samples
\citep[e.g.][]{2000AJ....120..165T, 2005ApJ...621..673T,
  2005ApJ...632..137N, 2006AJ....131.1288B, 2006astro.ph.10724J} as
well as the redshift evolution of luminosity and mass functions of
massive galaxies \citep[e.g.][]{2006MNRAS.372..537W,2006A&A...453L..29C,2007astro.ph..3276R}.

Until very recently, the predictions from semi-analytic models of
hierarchical galaxy formation were in severe conflict with the
observationally derived formation ages and timescales of early-type
galaxies. These models take the merger trees of dark matter haloes in
a $\Lambda$CDM universe either by following them in a cosmological
N-body simulation \citep{1999MNRAS.303..188K}) or generating them
using the extended Press-Schechter formalism
\citep{1991ApJ...379...52W}. They then calculate the more complex fate
of baryonic matter inside them with prescriptions for star formation
\citep{1998ApJ...498..541K}, radiative cooling of gas
\citep{1993ApJS...88..253S, 2001MNRAS.328..726S} and feedback from
supernovae \citep{1999MNRAS.303..188K}. Early semi-analytic models
such as that of \cite{1998MNRAS.294..705K} predicted that the most
massive galaxies in the universe harbour the youngest stellar
populations and assemble the last, contrary to what has been deduced
from observations. For a critical example,  these models struggle to match the observed chemical
abundance ratios and their trends with galaxy mass
\citep{1999MNRAS.306..655T, 2005MNRAS.363L..31N}.

The reason for this failure in the early models is that star formation
continues unchecked in massive haloes where supernova feedback is
insufficient \citep{1986ApJ...303...39D}.  \cite{2003ApJ...599...38B}
showed that in massive galaxies, an additional, more powerful heating
source than
SNe is needed to heat the cold gas that fuels star formation and
therefore stop it. This additional heating source is now generally
identified as feedback from active galactic nuclei (AGN) where jets
and outflows from the central supermassive black hole are presumed to
heat and expel most
residual gas via a wind thus preventing further star formation. 
This
\textit{AGN feedback} has been discussed by a number of authors
\citep{1997ApJ...487L.105C, 1998A&A...331L...1S, 2004MNRAS.347.1093B,
2005MNRAS.361..776S, 2005MNRAS.364.1337S} and is now part of many galaxy formation
models
\citep{2001MNRAS.324..757G, 2004ApJ...600..580G,
  2005Natur.435..629S,2005MNRAS.360...60K,2006MNRAS.365...11C,
  2006MNRAS.370..645B, 2006MNRAS.366..499D,2006MNRAS.370.1651C,
  2007MNRAS.377...63C,2007arXiv0704.3941C, 2005MNRAS.358L..16K}.
The phenomenological model of \citet{2006Natur.442..888S} is specifically motivated by
\textit{GALEX} observations of early-type
galaxies. \cite{2005Natur.433..604D} and
\cite{2005MNRAS.361..776S} have also investigated the effect of AGN
feedback in hydro simulations of mergers and have now placed them in
large-scale simulations probing cosmological volumes
\citep{2007arXiv0705.2269D, 2007arXiv0705.2238S}.
Note that alternatives have been suggested based on the processes of
cooling and shock-heating of baryonic matter in dark matter haloes
\citep{2007astro.ph..3435B,2007arXiv0704.2418K,2007arXiv0706.1246H}.

AGN feedback may come in two different modes at different epochs
during cosmic time. In an early phase at high redshifts the AGN may
truncate and quench star formation leading to the formation of the
dominant old population in massive galaxies. This so-called 'quasar
mode' \citep{2006MNRAS.365...11C} is most likely accompanied by vigourous
black hole growth and leads to very strong \textit{kinetic} feedback
that quenches
strong starbursts. The other face of AGN feedback can be thought of as a 'maintenance' mode. It turns on at more recent epochs such
that further significant episodes of star formation are prevented
\citep{2005MNRAS.362...25B, 2006MNRAS.368L..67B, 2005MNRAS.360L..20F,2006MNRAS.366..417F}. The
coupling between the black hole accretion and the gas
heating is poorly constrained by the observation. The latter mechanism
has only recently been included in semi-analytic galaxy formation
models.  While such prescriptions are
successful at curing some key issues, their major problem is that the
physics of this process is still poorly understood and essentially
unconstrained by observations. Compelling observational evidence for
the \textit{mere existence} of AGN feedback in early-type galaxies is still
missing.

The aim of this paper is to tackle the key question of the role of AGN
feedback observationally by looking for its signatures in early-type
galaxies at recent epochs in the nearby universe. Crucial for this
purpose is the compilation of a sample of early-type galaxies with a
selection method that does not bias against AGN and/or star
formation. The sample
further needs to be large, because the potentially short duration of
the AGN feedback phase makes such objects relatively rare. To this end
we have constructed a sample of nearly 16,000  early-type galaxies
from the Sloan Digital Sky Survey (SDSS) in the redshift range
$0.05\leq z\leq 0.1$.
These early-types are part of a sample comprising 50,000 early- and
late-type galaxies in that redshift range compiled in a project called
Morphologically Selected Ellipticals in SDSS (MOSES, Schawinski et
al., in preparation). The major novelty is the purely morphological
selection mechanism through visual inspection of \textit {all} 50,000
galaxies. Through emission line diagnostics we identify active
early-type galaxies with star formation activity and/or AGN. From the
stellar populations we derive star formation histories for these
objects, which allows us to explore the possible interaction between
the formation histories of the galaxies and their current star formation/AGN
activities.

The paper is organised as follows. In
Sections~\ref{sec:sample_selection} and \ref{sec:method}, we outline
the sample selection, present the emission line diagnostics, and the
stellar population analysis. We place the sample galaxies subdivided
in quiescent, AGN, and star forming objects on the colour-$\sigma$
relation and discuss first tentative conclusions in
Section~\ref{sec:redblue}. A more quantitative discussion is provided
in Section~\ref{sec:results} based on the two-component star formation
histories derived. Conclusions in the context of AGN feedback in
galaxy formation are discussed in Sections~\ref{sec:discussion}
and~\ref{sec:summary}.

Throughout this paper, we assume a standard $\Lambda$CDM cosmology
consistent with the WMAP results \citep{2003ApJS..148....1B}
with $\Omega_{m} = 0.3$, $\Omega_{\Lambda} = 0.7$ and $\mathrm{H_{0}} =
70~\mathrm{h_{70}~km^{-1}~s^{-1}}$. All magnitudes are in the AB
system.

\section{Data sample}
\label{sec:sample_selection}

\begin{figure}
\begin{center}

\includegraphics[angle=90, width=0.5\textwidth]{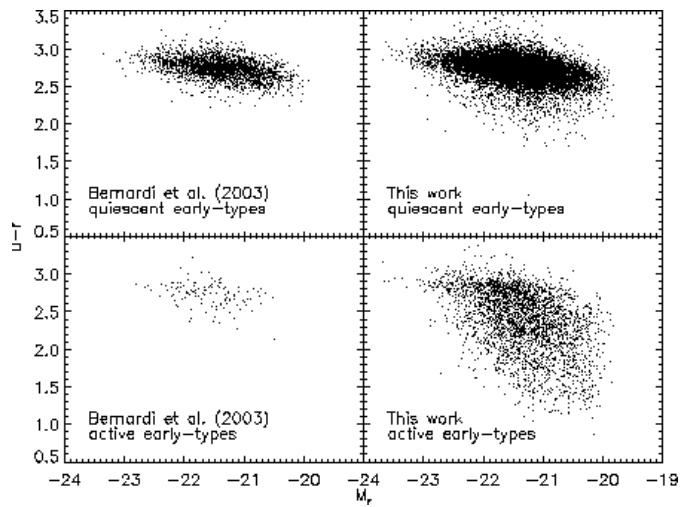}
\caption{Optical colour-magnitude relation for the early-type catalogue
  of \citet{2003AJ....125.1817B} and this work. Early-type galaxies
  are divided into 'quiescent' and 'active' depending on whether
  emission lines are detected. Our purely morphologically selected
  sample contains a significantly larger fraction of active
  early-type galaxies.}

\label{fig:bernardi_comparison}

\end{center}
\end{figure}

The sample utilised here is part of a project called MOSES:
\textbf{MO}rphologically \textbf{S}elected \textbf{E}llipticals in
\textbf{S}DSS. Key is that all 48,023 galaxies in this sample have been
visually inspected and classified into early- and late-type morphology
by eye, yielding a sample of 15,729 early-type galaxies.  A detailed description of the catalogue can be found in a companion paper (Schawinski et al., in preparation), and we refer the reader to this paper for any details. In the following, we provide a brief summary with focus on the aspects that are most relevant for the present work.

\subsection{Sample Selection}
We use the Sloan Digital Sky Survey DR4 as our main sample
\citep{2000AJ....120.1579Y, 2002AJ....123..485S,
2006ApJS..162...38A}. The \textit{SDSS} is a survey of the Northern
Sky providing us with photometry in the five filters \textit{u,g,r,i}
and \textit{z} \citep{1996AJ....111.1748F, 1998AJ....116.3040G}.  We
selected objects from the redshift slice $0.05 < z < 0.10$ with the
magnitude limit $r < 16.8$. These cuts are mostly motivated by the
limit out to which a visual inspection of galaxy morphology is
reliable (see \citealt{2006astro.ph..1036S}).

\subsection{Morphological Classification}

The selection criteria used to identify early-type galaxies are
crucial. Selection criteria based on galaxy properties such as
colour, spectral features or concentration index rather than
morphology are at risk to be biased against early-type galaxies
with residual star formation or AGN activity. As these are the objects
most relevant for the present study, we decided for a radical, purely
morphological selection.

The first catalogues of early-type galaxies in the \textit{SDSS} used for extensive study are those of
\cite{2001AJ....122.2267E} and \cite{2003AJ....125.1817B,
2003AJ....125.1849B, 2003AJ....125.1866B, 2003AJ....125.1882B}. The
\cite{2003AJ....125.1817B} catalogue is based primarily in the
likelihood of the surface-brightness profile resembling a de
Vaucouleurs law and a high concentration index. This is a very
efficient way of selecting large numbers of spheroids, but can fail to
reject spiral galaxies with prominent bulges. In particular, spirals
with face-on faint disks and edge-on disks can pass these criteria.
Furthermore, \cite{2003AJ....125.1817B} reject galaxies with a PCA
classification number atypical of red, passive early-types
\citep{1999AJ....117.2052C}. This means that early-types with strong
emission lines or that are dominated by young stellar populations are
rejected despite having early-type morphology. For an example, the
early-type galaxies discussed by \citet{2004ApJ...601L.127F} would not
be included. Recently, \citet{2007AJ....134..579F} have compiled an catalogue of morphological  early-type galaxies by visual inspection of 2,658 objects from the \textit{SDSS}.

For the MOSES sample we carried out manual inspection of multi-colour images from SDSS by eye \textit{for all 48,023 galaxies}, in order to avoid
any potential bias introduced by selecting by colours or spectral
features. We define as early-type galaxies all objects earlier than and
including S0 galaxies and reject galaxies from our sample if they show
distinct spiral arms or a disk. We \textit{include} in this selection
objects with clear tidal features or other signs of morphological
disturbance. In the paper presenting the MOSES catalogue (Schawinski et al., in preparation) we present a number of example images from
\textit{SDSS} to illustrate the effectiveness of our visual
inspection. Out of the 48,023 galaxies in our volume limited sample,
we identify 15,729 early-type galaxies. Figure
\ref{fig:bernardi_comparison} illustrateSs the difference between our
selection and that of \citet{2003AJ....125.1817B} in the
colour-$\sigma$ diagram. It can be seen clearly that our morphological
selection yields a significantly higher fraction of early-type
galaxies with emission lines and/or blue colours. Less than 5\% of the
early-type galaxies in the catalogue of \citet{2003AJ....125.1817B}
show emission lines, while over 18\% of ours do.

\subsection{Photometry}
We retrieved \textit{SDSS} model magnitudes (\texttt{modelMag}) for
all five filters \textit{u,g,r,i} and \textit{z} from the
\textit{SDSS} DR4 database. In order to improve the wavelength
coverage of the galaxies in our sample, we match them to a number of
other surveys, namely \textit{GALEX}, \textit{2MASS}, and
\textit{SWIRE}.

\subsubsection{Matching to GALEX}
The \textbf{Galaxy Evolution Explorer} (\textit{GALEX};
\citealt{2005ApJ...619L...1M}) is conducting several surveys of the
ultraviolet sky in two filters: far-UV (\textit{FUV}; 1344-1786\AA)
and near-UV (\textit{NUV}; 1771-2831\AA). The \textit{Medium Imaging
Survey} from the General Release 3 (MIS/GR3) is covering a portion of
the sky coinciding with the \textit{SDSS} to a limit of 23 AB in
\textit{NUV} and \textit{FUV}. We match our galaxies to MIS within
$4\arcsec$ to find counterparts. GR3 only covers a fraction of
\textit{SDSS} DR4 and so this matching gives us detections for 13.4\%
of galaxies.

\begin{figure*}
\begin{center}

\includegraphics[angle=90, width=\textwidth]{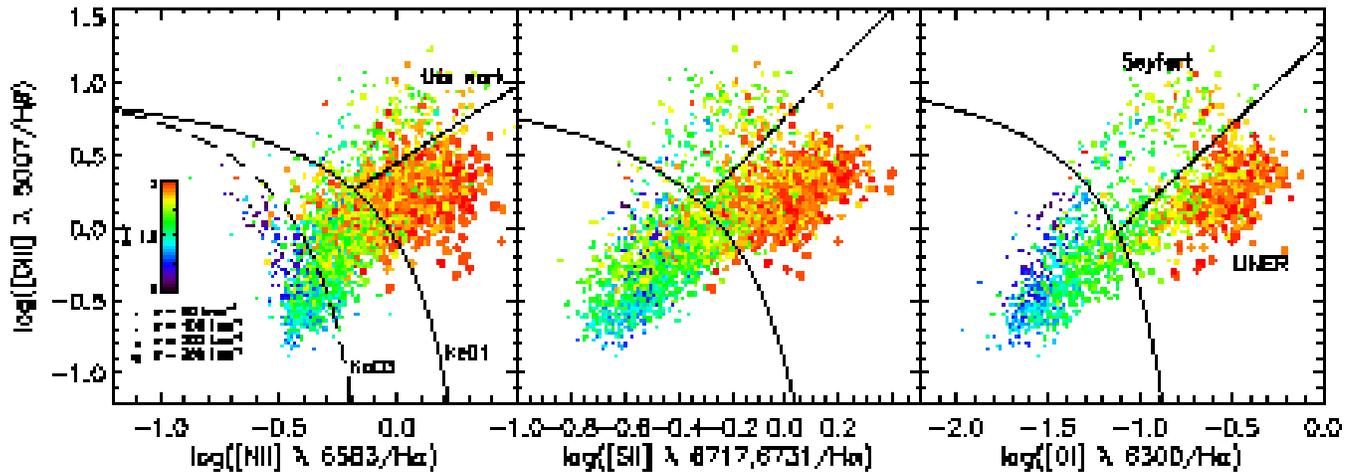}
\caption{The BPT line diagnostic diagrams for the early-type galaxies
of the MOSES sample. Each galaxy is coloured by its optical u-r colour (see
colour bar in left-hand panel). Point sizes scale with the galaxy
velocity dispersion as an indicator of mass (see legend in left-hand
panel). In each diagram, we indicate the demarcation lines used in our
classification scheme. The dashed line in the [NII]/H$\alpha$ diagram
(left-hand panel) is the empirical star formation line of
\citet{2003MNRAS.346.1055K} (labelled Ka03), while the solid curve (in
all three panels) is the theoretical maximum starburst model from
\citet{2001ApJ...556..121K} (labelled Ke01). The galaxies in between
these two lines are SF-AGN composites or 'transition region
objects'. Galaxies below the Ka03 line are dominated by star
formation. The division line between Seyferts and LINERs is shown as
the straight line following \citet{2006MNRAS.372..961K} for the middle
and right-hand panel and our own definition (see text) for the
left-hand panel.\label{fig:bpt_plots}}

\end{center}
\end{figure*}

\subsubsection{Matching to 2MASS}
The \textbf{Two Micron All Sky Survey} (2MASS;
\citet{2006AJ....131.1163S}) provides us with \textit{J, H} and $K_s$
band imaging for our galaxies selected in \textit{SDSS}. We match our
galaxies to the \textit{Extended Source Catalogue} (XSC) within
$4\arcsec$, giving us counterparts for 87.1\%. The \textit{2MASS}
photometry is done in the Vega system, so in order to keep consistency
with the rest of the MOSES sample, we convert the magnitudes from Vega to AB
following \citet{2003AJ....125.2348B, 2005AJ....129.2562B}.

\subsubsection{Matching to Spitzer-SWIRE}
The \textbf{Spitzer Wide Area Extra-galactic Survey} (SWIRE;
\citealt{2003PASP..115..897L}) is one of the Spitzer Legacy Surveys
and imaged nearly 50 square degrees using both the IRAC ($3.6, 4.5,
5.8, 8.0\mathrm{\mu m}$) and MIPS ($24, 70, 160\mathrm{\mu m}$)
instruments. We cross-match the MOSES sample to the public source catalogues
for ELAIS N1, N2 and Lockman Hole fields. Only a small fraction of our
sample (0.3\%) has counterparts at $3.6\mathrm{\mu m}$.

\subsection{Spectroscopy}
\label{sec:spectroscopy} 

\textit{SDSS} has performed follow-up spectroscopy using $3\arcsec$
fibres yielding a reasonably complete sample to $r < 17.77$
\citep{2003AJ....125.2276B}. The spectra cover the range of 3800 to
9200\AA~at a resolving power of 1800 and for the MOSES sample have a typical
signal-to-noise of about 30. We retrieve the \textit{SDSS} optical
spectra for all our galaxies.

\subsubsection{Gas and stellar kinematics}
We measure the kinematics of the stars and the emission-line fluxes of
our galaxies from their SDSS spectra following the methods of
\citet{2004PASP..116..138C} and \citet{2006MNRAS.366.1151S},
respectively\footnote{We made use of the corresponding \texttt{ppxf}
and \texttt{Gandalf} IDL codes adapted for dealing with SDSS
data. Both softwares can be retrieved at
http://www.strw.leidenuniv.nl/sauron/}.
We separate the contribution of the stellar continuum and of the
ionised-gas emission to the observed galaxy spectrum by fitting
\textit{simultaneously} stellar population templates
\citep[from][]{2004ApJ...613..898T} and Gaussian emission-line
templates to the data.
In these fits we account also for the impact of diffuse dust in the
galaxy of dust in the emission-line regions, which allows to obtain a
decrement on the strength of the Balmer lines that is at least what
expected by recombination theory.
From the fit to the stellar continuum and absorption features, we
measure the line-of-sight velocity dispersions. From subtraction of
the emission-line spectrum from the observed one, we get the clean
absorption line spectrum free from emission line contamination. We
then use this cleaned spectrum to measure the stellar absorption
indices. 
The physical constraints on the emission from high-order Balmer lines
ensures the strength of the corresponding absorption features is
correctly estimated, which is crucial for constraining
the ages of stellar populations.

\subsubsection{Lick absorption line indices}
On each spectrum we measure the 25 standard Lick absorption line
indices \citep{1994ApJS...94..687W, 1997ApJS..111..377W} following the
most recent index definitions of \cite{1998ApJS..116....1T}. For this
purpose the spectral resolution is reduced to the wavelength-dependent
Lick resolution \citep{1997ApJS..111..377W}. The measurements are then
corrected for velocity dispersion broadening. The correction factor is
evaluated using the best fitting stellar template and velocity
dispersion obtained previously. Errors are determined by Monte Carlo
simulations on each spectrum individually based on the signal-to-noise
ratios provided by the \textit{SDSS}. Possible discrepancies between
the flux-calibrated \textit{SDSS} spectral system and the Lick system are
negligible, as shown in a detailed analysis of Lick standard stars
observed with \textit{SDSS} \citep{Carson07}.

\subsection{Emission line diagnostics}
\label{sec:eml_diag}

\begin{figure*}
\begin{center}

\includegraphics[angle=90, width=0.19\textwidth]{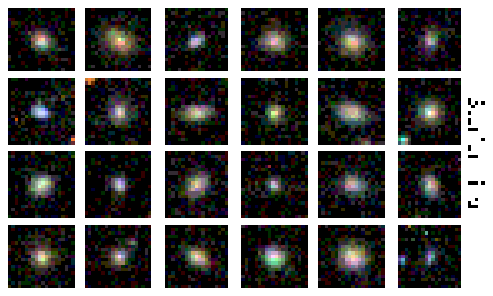}
\includegraphics[angle=90, width=0.19\textwidth]{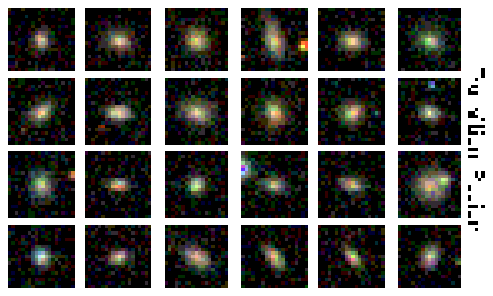}
\includegraphics[angle=90, width=0.19\textwidth]{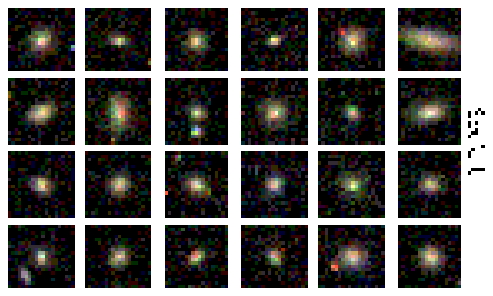}
\includegraphics[angle=90, width=0.19\textwidth]{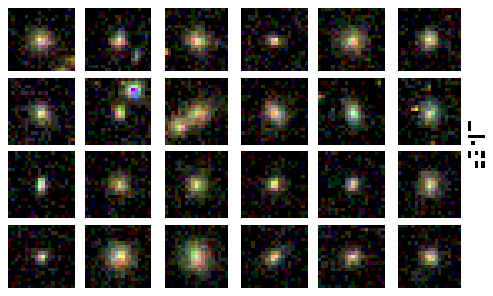}
\includegraphics[angle=90, width=0.19\textwidth]{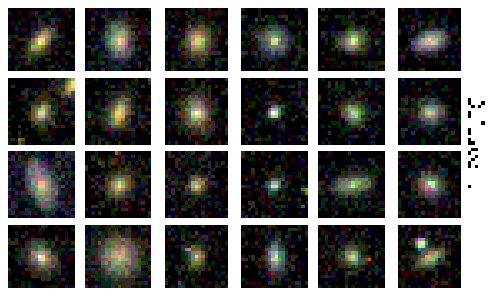}

\caption{A sample of morphological early-typegalaxies from the MOSES
  sample sorted by their emission line classification. The
  cutouts are 40$h^{-1}$ kpc by 40$h^{-1}$ kpc in size.\label{fig:sfe_example}}

\end{center}
\end{figure*}

Emission line diagnostic diagrams are a powerful way to probe the
nature of the dominant ionising source in galaxies \citep*[][hereafter
BPT]{1981PASP...93....5B}. They allow the separation of galaxies into
those dominated by ongoing star formation and non-stellar processes;
and with sufficient information can further split those into Seyfert
AGN and LINERs\footnote{LINER: low ionisation nuclear emission line
region}. The diagrams also contain a transition region, where the
emission lines indicate a blend of star formation and AGN activity.

We use the four optical line ratios [OIII] $\lambda$5007/H$\beta$,
[NII] $\lambda$6583/H$\alpha$, [SII] $\lambda$6583/H$\alpha$ and [OI]
$\lambda$6583/H$\alpha$. We follow the signal-to-noise criterion of
\citet{2003MNRAS.346.1055K} and classify all galaxies that have a S/N
$>$ 3 detection of H$\alpha$, H$\beta$, [OIII] $\lambda$5007 and [NII]
$\lambda$6583. The two low ionisation species [SII] and [OI] are used
when detected, as they are usually weaker than the other four.

\subsubsection{Diagnostic diagrams}
The corresponding three diagnostic diagrams are shown in
Fig.~\ref{fig:bpt_plots}. Each galaxy is coloured by its optical $u-r$
colour. Point sizes scale with the galaxy velocity dispersion as an
indicator of mass. In each diagram, we indicate the demarcation lines
used in our classification scheme. In the left-hand column, we show
the [NII]/H$\alpha$ diagram used to separate star forming objects
(blue) by means of the demarkation line by \citet[][dashed
line]{2003MNRAS.346.1055K}. We verified that the somewhat more
restrictive separation between star forming and AGN suggested by
\citet{2006MNRAS.371..972S} does not alter the
results of this work. The remaining objects are divided in composite
Transition Region objects (purple) and AGN using the theoretical
maximum starburst model from \citet[][solid
lines]{2001ApJ...556..121K}.

The AGN are then further sub-classified into Seyferts (green) and
LINERs (red) by the straight solid lines. The more indicative low
ionisation species (middle and right-hand panel) are used where
possible.  The best separation of Seyferts and LINERs is achieved
using [OI] (right-hand panel of Fig.~\ref{fig:bpt_plots}), which is
used wherever [OI] is detected with S/N $>$ 3. The [OI] line is
generally weaker then any of the other emission lines used, so where
it is not available, we use [SII]. To separate the two types of AGN,
we use the lines defined by \citet{2006MNRAS.372..961K}. If both these
lines are not detected we adopt the original [NII] diagram for AGN
classification. Based on the location of those galaxies classified by
[OI] or [SII] on the [NII] diagram, we define a new empirical
Seyfert-LINER separation:

 \begin{equation}
  \mathrm{log([OIII]/H\beta) = 1.05~log([NII]/H\alpha) + 0.45}
 \label{eqno1}
 \end{equation}

\subsubsection{Classification results}
Following the emission line classification we divide our early-type
galaxy sample into five classes.

\begin{enumerate}
\item Quiescent (orange)
\item Starforming (blue)
\item Transition Region (purple)
\item Seyfert (green)
\item LINER (red)
\end{enumerate}

Most objects in the quiescent early-type galaxy class are virtually
free of emission lines. A small sub-fraction exhibits weak [OIII]
emission, but lack the other emission lines necessary for the BPT
classification. We verified that a further separation of this subclass
does not affect the results of this work. The quiescent objects make
up the bulk of the early-type population (82 per cent). The early-type
galaxies with significant emission lines (S/N $>$ 3 in the four main
BPT lines) represent the remaining fifth of the early-type sample. We
call these active early-type galaxies, and separate them into star
forming, transition objects, Seyferts and LINERs. The results of this
classification are summarised in Table \ref{tab:class_results}.
For further details we refer the reader to the paper describing the
MOSES catalogue (Schawinski et al, in preparation). We show some
example images of morphological early-type galaxies in Figure
\ref{fig:sfe_example}.

\begin{table}
\caption{Emission Line Classification Results}
\label{tab:class_results}
\begin{tabular}{@{}lrrl}
\hline
\hline
Classification & \multicolumn{1}{c}{Number} & \multicolumn{1}{c}{Fraction}  & Colour\\
 &  &  Galaxies & \\
\hline
Early-type galaxies & 15729 & 100\% &\\
\hline
Quiescent & 12828& 81.5\%& orange\\
Starforming & 671 & 4.3\%& blue\\
Transition Region & 1087& 6.9\%& purple\\
Seyfert & 242 & 1.5\%& green\\
LINER & 901& 5.7\%&red\\
\hline
\end{tabular}
\end{table}

\subsubsection{Star formation rates}
For those galaxies classified as starforming, we derive star formation
rates from the extinction-corrected H$\alpha$ line luminosity
according to the conversion of \cite{1998ApJ...498..541K}. All star
formation rates for our galaxies quoted in this paper are based on
H$\alpha$ unless otherwise noted. We cannot derive reliable star
formation rates for any of the other emission line classes, as an
unknown fraction of the H$\alpha$ emission line is of non-stellar
origin.

\section{Stellar population analysis}
\label{sec:method}

The spectrum of a galaxy contains a record of its star formation
history over the age of the universe. Its stellar populations are the
combination of all the episodes of star formation that a galaxy and
its progenitors underwent. Reconstructing this star formation history
or at least recovering it partially is a challenging task that
requires excellent observational data, stellar population models for
the interpretation and a statistically sound method for the recovery.

For the problem at hand in this paper, we turn to the
parametric approach. It has been used with great success by
\cite{2000ApJ...541L..37F} and by \cite{2006astro.ph..1029K} to
use the immense discriminating power of restframe near- and
far-UV photometry to test whether individual early-type
galaxies have undergone episodes of recent residual star
formation or not. 

\subsection{Two-Component star formation histories}
\begin{figure}
\begin{center}

\includegraphics[angle=90, width=0.5\textwidth]{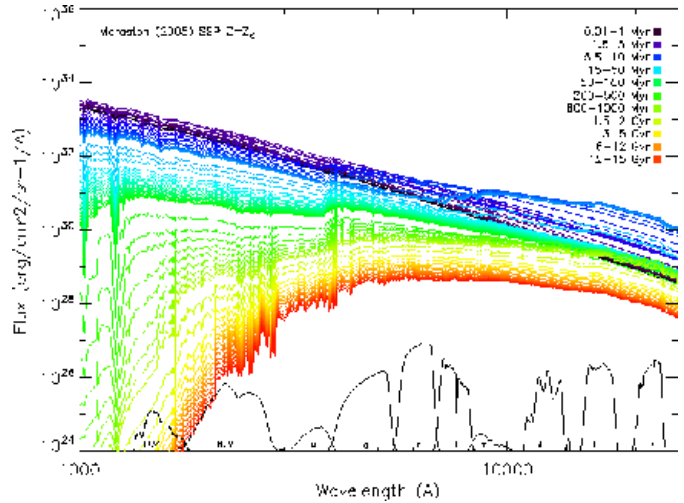}
\caption{The pace of stellar evolution: we show the SED of a
  solar metallicity M05 SSP from 0.01 Myr to 15
  Gyr. The different ages are colour-coded and we indicate the filters
  we use by their filter transmission curves of the data we use in
  this paper at the bottom. The
  broadband colours of a galaxy will always be dominated by the
  youngest stellar population formed in the most recent episode of
  star formation. Any older population will be small by comparison
  even with respect to a relatively small young mass-fraction. This
  feature of stellar evolution allows us to effectively split the
  star formation history into two components until about 1.5-2 Gyr
  when the pace of stellar evolution slows down and populations of
  different ages up to 15 Gyr become similar and the differences are
  on the order of the effects of dust and metallicity.
\label{fig:age_evolution}}

\end{center}
\end{figure}

The aim of this work is the mapping of current star formation/AGN
activities in galaxies with their star formation histories. Of
particular relevance for us is the epoch and the strength (hence mass
fraction) of the last significant starburst that occurred in a galaxy. 
Since the most recent episode of star formation dominates
the spectrum at all wavelengths, it is the easiest to recover (see
Figure \ref{fig:age_evolution}). Following the approach of \citet{2006astro.ph..1029K}, we
therefore fit two-component star formation histories to the galaxy
data. 

The old component is modelled as an instantaneous starburst of
variable age and metallicity, and represents the formation of the bulk stellar
population of the galaxy before the most recent episode of star
formation. We allow this old component to vary between a $t_{o}$ of 1
and 15 Gyrs. By marginalising over this parameter later on, we are
summing all possible linear combinations of single starbursts and so
take into account all possible star formation histories before the
most recent starburst.

On top of this, we add a young component of a certain age $t_{y}$ and
mass fraction $f_{y}$. It turns out that this young burst is poorly
modelled by an instantaneous starburst, so that we choose to represent
it by an exponentially declining star formation rate with an e-folding
time of 100 Myr. The choice of 100 Myr is motivated by a test on a
small sub-sample, where
we left the e-folding time $\tau$ as an additional free parameter. In
this test, we recovered a typical $\tau$ of $\sim$100 Myr for
early-type galaxies with young components, so we set $\tau$ to 100 Myr
\textit{a posteriori}.  The parameter $t_{y}$ stands for the look-back
time to the beginning of the starburst, $f_{y}$ is the total mass
fraction formed in the recent burst extrapolating the 100 Myr
e-folding star formation history to infinity. Since virtually
all ages we recover for the starburst population are at least
$\sim$100 Myr, the difference between the mass actually formed by
today and the total mass that will form is small.

This parametric quantification of the age and mass fraction of the
secondary population allows us to derive a \textit{recent} star
formation history against the background of any possible star
formation histories in the distant past, represented by a simple
stellar population (SSP), whose effect we can account for by
marginalisation. This approach is different from purely non-parametric
approaches in the literature that make no \textit{a priori}
assumptions on the star formation history to be recovered
\citep[e.g.][]{2003MNRAS.343.1145P,2006astro.ph..8531P}.

\begin{figure*}
\begin{center}

\includegraphics[angle=90,width=0.93\textwidth]{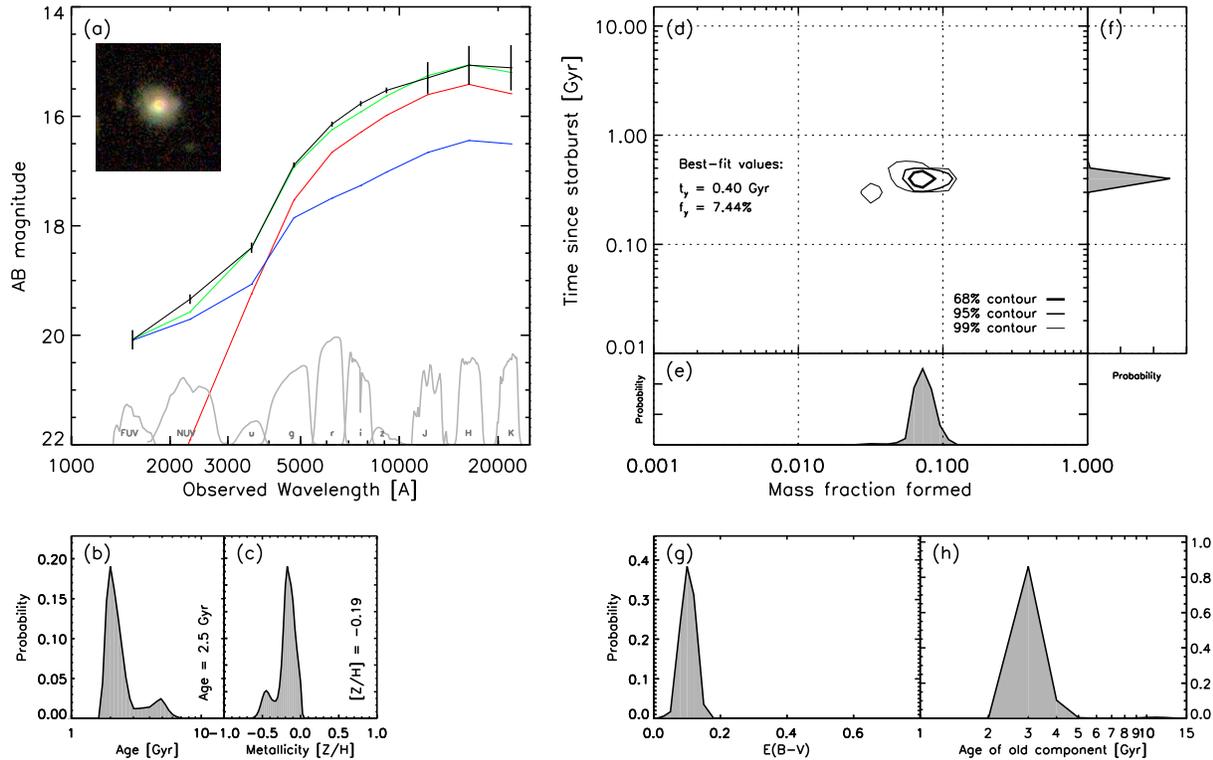}
\caption{An example of a galaxy where we can resolve and quantify the last
  significant episode of star formation. This object is a
  Starforming early-type galaxy. For a detailed description,
  see section \ref{sec:example_gals}.\label{fig:young_example}}

\end{center}
\end{figure*}

\begin{figure*}
\begin{center}

\includegraphics[angle=90,width=0.93\textwidth]{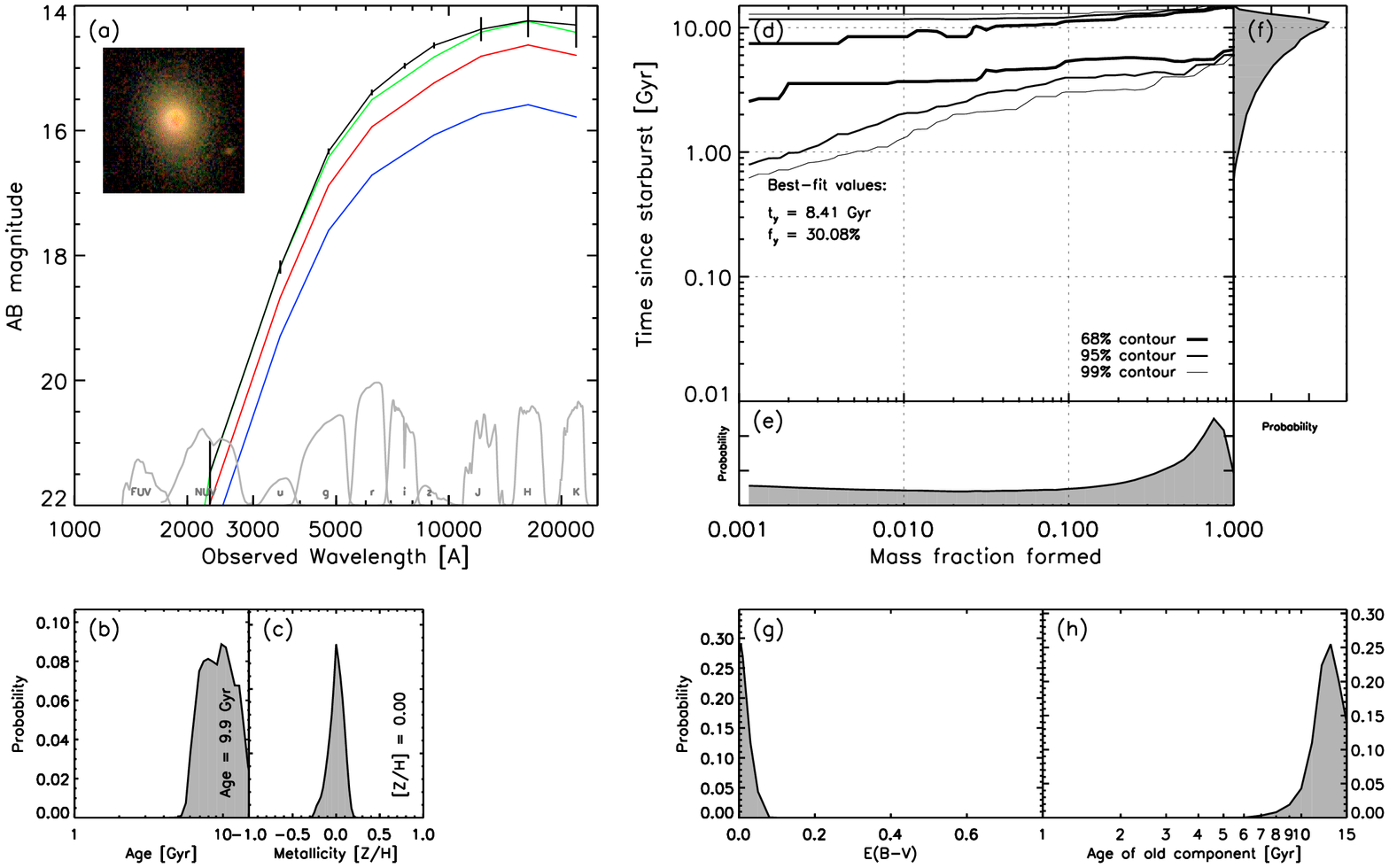}
\caption{An example of a galaxy where we cannot quantify the first
  significant episode of star formation, but can place strong limits
  on the recent star formation history. For a detailed description,
  see section \ref{sec:example_gals}.\label{fig:old_example}}

\end{center}
\end{figure*}

\subsection{Internal metallicity distribution}
Besides assigning an average metallicity to each model, we
additionally consider the internal distribution of metallicities
necessarily present in each galaxy. This is important, as a tail of
low metallicity stars can have a significant effect in particular on
the ultraviolet part of the spectrum \citep{2000ApJ...541..126M}. 
We construct an analytic form of the typical metallicity distribution
from observations of the bulge of M31 and the closest early-type
galaxy NGC~5128 where individual stars can be resolved
\citep{2005AJ....130.1627S,1999AJ....117..855H,2000AJ....120.2423H,
2002AJ....123.3108H}. This metallicity distribution is modelled with
two Gaussians. The bulk metallicity is a Gaussian with a mean
metallicity $\mu_Z$ and a spread of $\sigma$ = 0.22 dex, while the
tail is parametrised as a Gaussian shifted by 0.75 dex to lower
metallicity with the much larger spread of $\sigma$ = 1.5 dex. Both
components are such metallicity composites, and we vary $\mu_Z$ to
cover a mean mass-weighted metallicity $Z(\mu_Z)$ ranging
between $\sim 1/200 Z_{\odot}$ to $\sim 3.5 Z_{\odot}$. 

\subsection{Dust extinction}
Internal dust extinction causes a degeneracy with age
and mass that needs to be taken into account. To describe dust
attenuation we adopt the extinction law by
\citet{2000ApJ...533..682C}. This choice is mainly motivated by the
fact that the active early-type galaxies in our sample are undergoing
significant episodes of star formation, so that an extinction law
derived from a sample of starforming galaxies seems reasonable. In
separate runs we have tested a number of other extinction laws (Milky
Way \citep{1976asqu.book.....A}, LMC \citep{1986AJ.....92.1068F}, SMC
\citep{1984A&A...132..389P, 1985A&A...149..330B} and verified that in
general, the \citet{2000ApJ...533..682C} law provides the best
fits. The dust parameter E(B-V) is varied between 0.0 and 0.8.

\subsection{Combining photometry and spectroscopy}
A crucial novelty of our approach is the combination of photometric
and spectroscopic data, which increases the power of breaking the
degeneracies between age, mass, metallicity and dust content. The
reason why this helps is that the degeneracies are different because
of the dissimilar \textit{non-linear} responses of broadband colours
and stellar absorption indices to these parameters. Like
cosmologists who combine the very degenerate contours of one
experiment with another to find a well-constrained set of parameters in
the intersection, we harness two very different observables
that couple to the fundamental physical parameters in very
different fashions.

It should be noted that,
absorption line indices are not significantly affected by dust
\citep{2005ApJ...623..795M}. A further cornerstone in
our fitting algorithm for the identification of intermediate-age
populations is the inclusion of near-IR photometry and stellar
population models \citep{2005MNRAS.362..799M} that adequately describe the
thermally pulsing asymptotic giant branch (TP-AGB) phase. This is
crucial as the TP-AGB provides most of the near-IR light of a stellar
population between 0.5 to 2 Gyr \citep{1998MNRAS.300..872M}.

In a nutshell, we perform a comparison of the observed photometric SED
(of as many as are available of \textit{FUV, NUV, u, g, r, i, z, J, H,
  K}) to a library of SEDs of over five million two-component star
formation histories from \cite{2005MNRAS.362..799M} and compute the
appropriate model SED parameterised by the five parameters described above
($t_{y}$, $f_{y}$, $t_{o}$, $\mu_{z}$ and $E(B-V)$) compute the
associated $\chi^2$ statistic to obtain a probability distribution
function for those five parameters. An intrinsic error for our photometry
across all filters of 0.015 mag is adopted to deal with the
uncertainties in the absolute photometric calibration of \textit{SDSS}

Simultaneously, we compute the associated  luminosity-weighted SSP
ages and metallicities for each model. We derive
luminosity-weighted SSP ages and metallicities by $\chi^2$ fitting the
\citet{2003MNRAS.339..897T,2004MNRAS.351L..19T} models to the 25 Lick
absorption line indices and compute a second probability distribution
through the comparison of these quantities with the $V$-luminosity
weighted ages and metallicities of the photometric comparison. These two
probability distributions from the SED and absorption line fitting
are then convolved to the final one.

In order to estimate the parameters of interest, we marginalise the
resulting 5-dimensional probability distribution function to obtain
the two-dimensional distribution in the recent starburst age vs.\
mass-fraction plane ($t_y$-$f_y$). We integrate over this surface to
obtain confidence levels constraining the age and mass-fraction of the
last significant episode of star formation in the galaxy. These
probability distributions are stacked to get typical values for the
sub-populations.

\subsection{Example galaxies}
\label{sec:example_gals}
In this section we present two example fits to galaxies with very
different properties and star formation histories. The first galaxy is
an active early-type in the transition region of the emission
line classification which suggest that the object is dominated both by
ongoing star formation and an AGN. Its H$\alpha$ star formation rate
is 2.1 $M_{\odot}yr^{-1}$ and its
UV-optical colours place it squarely into the blue cloud ($u-r$ = 2.3,
$NUV-r$ = 3.2).  The second galaxy is a typical passive
quiescent early-type galaxy with no detected emission
lines. Its UV-optical colours place it on the red sequence ($u-r$ = 2.8,
$NUV-r$ = 6.1). The availability of \textit{GALEX} \textit{NUV} and \textit{FUV}
photometry generally improves the confidence contours, but does not
shift the best fit.

The fit results are summarised in Figs.~\ref{fig:young_example}
and~\ref{fig:old_example}. In panel (a) we show the fit to the
photometry: in black are the observed broadband colours with the
errors magnified by a factor of 3 to make them visible. To guide the
eye, we show the filter transmission curves of the various filters
used from \textit{GALEX} far-UV to \textit{2MASS} $K$-band. The best-fit
model is shown in green and the separate old and young components are
red and blue respectively. We also show the \textit{SDSS} colour
images.

In panels (b) and (c) we show the probability distributions for the
luminosity-weighted age and metallicity derived from the absorption
line indices. The final result in the $t_y$-$f_y$ plane is shown in
panel (d). This plot quantifies strength and epoch of most recent
episode of star formation. In panels (e) and (f), we show the
projected, projected probability distributions for the parameters
$t_y$ and $f_y$. In the main analysis we will stack these
distributions for various classes of galaxies. In panels (g) and (h),
we show dust extinction $E(B-V)$ and the age of the old component.

The two example galaxies show very different star formation
histories. The active galaxy has a very young luminosity-weighted age,
which is well consistent with the increased UV flux. The two-component
SED fit yields an old component of 3 Gyr and places the most recent
starburst at an age of 400 Myr with a mass fraction of 7 per cent.
Its dust content is relatively high ($E(B-V)=0.1$) in agreement with
the current star formation activity. The low luminosity-weighted age
of the 'old' component implies that this object had several
rejuvenation events in the past.

The emission line free galaxy, instead, has an old
luminosity-weighted age derived from the spectroscopic data. The
two-component fit yields basically two old components with ages around
10 Gyr. We can conclude that this galaxy clearly has not undergone a
significant starburst in the recent past, while the details of its
star formation history in the more distant past are not
resolved. The 99\% confidence intervals exclude any sort of significat
star formation activity in the recent past. The
very low dust attenuation value fits well into this picture.

We also note that while only a fraction of the galaxies in our sample
have \textit{GALEX}, their fitted star formation histories do not
differ significantly from those without.

\begin{figure*}
\begin{center}

\includegraphics[angle=90, width=\textwidth]{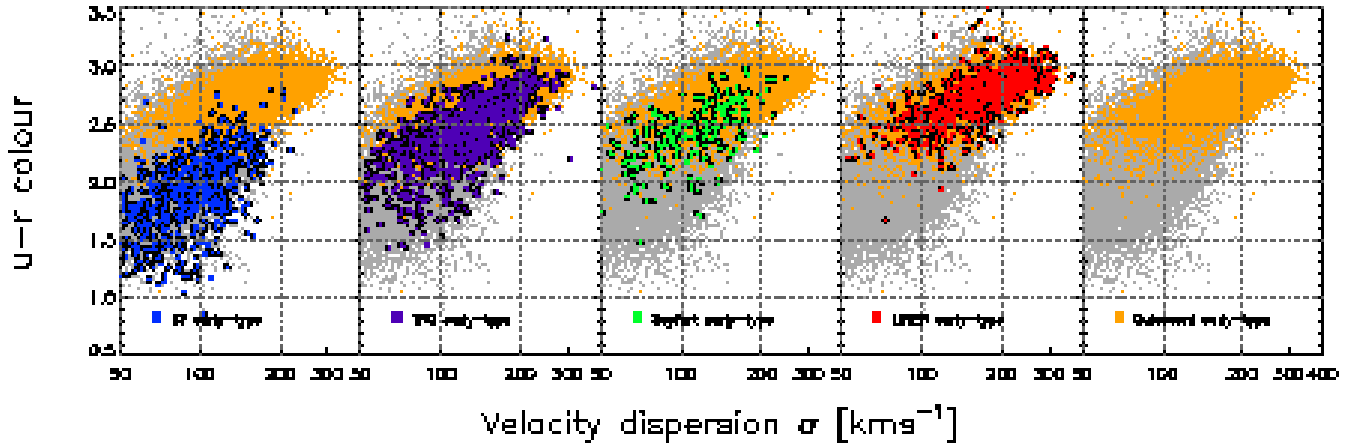}
\caption{The colour-sigma relation for our sample. In each panel,
morphological late-types are grey, quiescent early-types are orange
and the various active early-types are blue (Starforming), purple
(transition region), green (Seyfert) and red (LINER). The mass
tracer is the stellar velocity dispersion. \label{fig:ur_csigma}}

\end{center}
\end{figure*}

\section{Red Sequence and Blue Cloud}
\label{sec:redblue}
In this Section, we discuss the position of the various galaxy classes
defined through the emission line diagnostics in the colour-mass
relation. This simple comparison allows us to draw first tentative
conclusions that are independent of our stellar population analysis
method. Quantitative results of the detailed analysis will be
presented in the subsequent section.

\subsection{Active galaxies in the colour-mass relationship}
In Fig.~\ref{fig:ur_csigma} we show \textit{u-r} colour as a function
of galaxy velocity dispersion $\sigma$ for the galaxies in our sample.
The morphological late-type galaxies are shown as grey
points, the early-type galaxies are coloured points subdivided in the
various emission line classes: quiescent (i.e.\ no emission lines,
orange), LINER (red), Seyfert (green), Transition (purple), and star
forming (blue). The quiescent early-type galaxies (see most right-hand
panel) form the well-known red sequence, while the late-type galaxies
in the sample define the so-called blue cloud (grey points).

The first key result of this work is that the active galaxies are
offset from the red sequence and exhibit
well-defined patterns across the colour-$\sigma$ diagram. Starforming
objects inhabit the blue cloud. While this may seem like an obvious
result, note that the broadband SED and the emission lines probe very
diffierent physical components of a galaxy. While the former traces
the stellar populations and therefore the past, fossil record, the
emission lines provide
information on the {\em current} state of the {\em ionised gas}
in the galaxy. It is very reassuring that objects with ongoing star
formation show significant rejuvenation also in their stellar
populations.

More importantly, those objects not dominated by star formation but by
AGN activity, Seyferts and LINERs, are located considerably closer to
and almost on the red sequence. Transition objects, instead, which are
composites of star formation and AGN activity are found right
between these two extremes. There is a sequence between the blue cloud
and the red sequence from star forming via transition region and AGN
to quiescence. In more general terms, the presence of AGN in a
\textit{green valley} between the blue cloud and the red sequence has
been discussed in other recent work in the literature
\citep{2005astro.ph..6044F,
  2006astro.ph..9436K,2007astro.ph..3281M,2007ApJ...660L..11N,
  2007arXiv0704.3611S, 2007arXiv0706.3113W,2007arXiv0707.1523G}.

\subsection{Mass dependence}
This tentative sequence is convolved with a clear mass dependence of
the emission line classification. The blue star forming objects are
found at the low-mass end, while the majority of early-type galaxies
classified as LINERs lie at the high-mass end on the red
sequence. Transition objects and Seyferts reside between these two
extremes also in their galaxy mass distributions.

\begin{figure}
\begin{center}

\includegraphics[angle=90, width=0.5\textwidth]{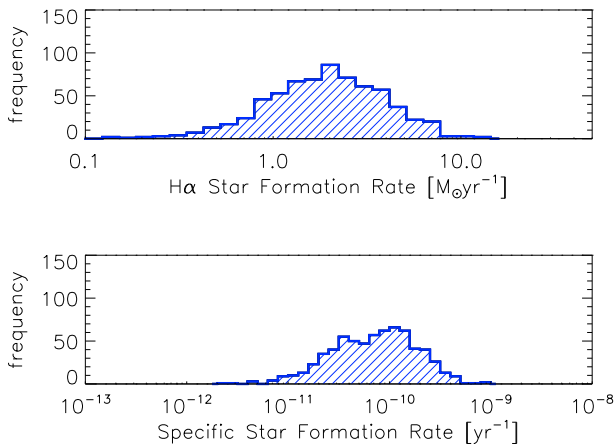}
\caption{Histogram of the star formation rate and specific
  star formation rate of the Starforming early-type galaxies. The
  star formation rate is based on the luminosity of the H$\alpha$ line
  and is converted to a rate using the relation of
  \citet{1998ApJ...498..541K}. \label{fig:sfr_hist}}

\end{center}
\end{figure}

The presence of a significant fraction of low-mass early-type galaxies
showing signs of ongoing star formation of a few $M_{\odot}$ per year
(see Fig.~\ref{fig:sfr_hist}) is a striking result, which is mostly
the merit of the pure morphological selection strategy. Still, this
result is not surprising.  The \textit{GALEX} UV satellite has found
evidence for minor recent star formation in early-type galaxies
\citep{2005ApJ...619L.111Y}, and the \textit{SAURON} survey of 48 nearby
elliptical and lenticular galaxies has shown that many of them have
(often extended) emission lines and signs of recent star formation as
well as CO emissions from molecular gas and atomic hydrogen indicating
sometimes substantial reservoirs of gas \citep{2006MNRAS.366.1151S,
2006MNRAS.371..157M, 2007astro.ph..3557C}.

In general, low-mass early-type galaxies are well-known to show young
stellar ages, emission lines and even ongoing star formation. This
downsizing of galaxy formation, wherein the most massive galaxies
harbour the oldest stellar populations is now generally
accepted. Today's early-type galaxies in particular show a clear
correlation between mass and age
\citep{2003AJ....125.2891C, 2005ApJ...621..673T, 2005ApJ...632..137N,
  2006AJ....131.1288B, 2006astro.ph.10724J}.

Fig.~\ref{fig:ur_csigma} shows that our sample comprises a large,
statistical sub-sample of active early-type galaxies in the
intermediate-mass and low-mass regimes ($\sigma\la 200~\rm{kms^{-1}}$)
allowing us to study the interplay between star formation, nuclear
activity, and quiescence in early-type galaxies. We note that many of
the high-mass transition region objects are most likely misclassified
LINERs. They do not have any starforming or Seyfert counterparts at
the same mass.

\subsection{AGN feedback in action?}
\label{sec:timesequence}
At low and intermediate masses we can identify a sequence at a given
mass from star formation via transition region and AGN activity to
quiescence, which is echoed in a sequence from blue to red stellar
populations. Taking \textit{u-r} colour as a proxy for the stellar age
and hence using the stars as cosmic clock, this is highly suggestive
of an evolutionary sequence. We are observing the act of galaxy
transformation from the star forming blue cloud via the green valley
to the passive red sequence \textit{in action}. 

We interpret the deviation from the red sequence as rejuvenation due to
a recent episode of star formation. On their way back to the red
sequence, galaxies appear to pass through a phase of significant nuclear activity, as
transition region objects and Seyferts have colours between the blue
star forming and the red sequence objects. This indicates that
AGN may have a decisive role to play in this transformation
process. Fig.~\ref{fig:ur_csigma} suggests that residual star
formation and nuclear activity are connected in early-type
galaxies. It is possible that the AGN actually suppress the star
formation activity, and we are witnessing AGN feedback in action. Our
sample represents the ideal laboratory to probe these processes.

It should be noticed that the massive early-type galaxies in our
sample do not appear to be part of such transformation process. At
the high-mass end only quiescent and LINER objects are found.

\begin{figure*}
\begin{center}

\includegraphics[angle=90, width=\textwidth]{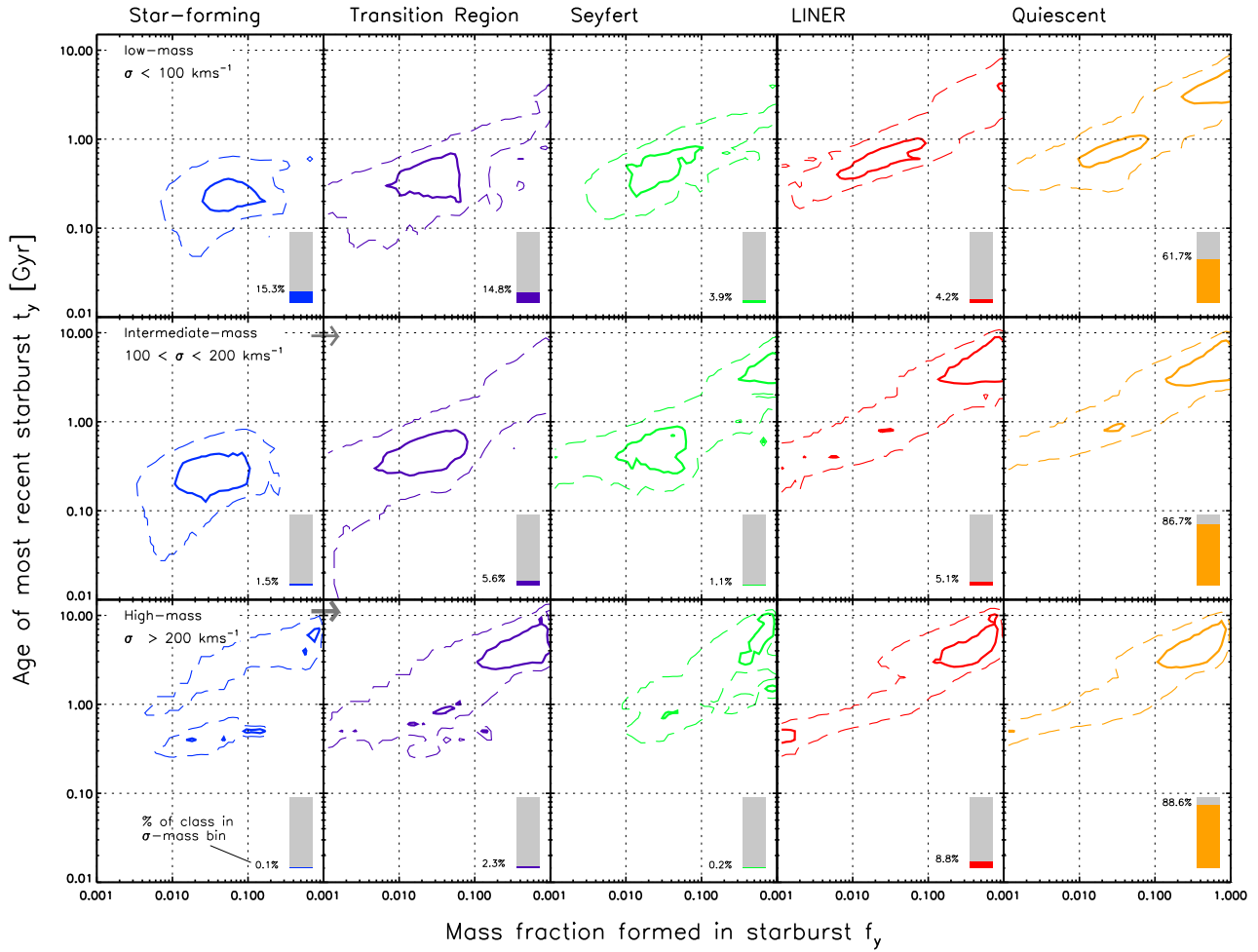}
\caption{The recent star formation histories of early-type galaxies
by mass and emission line class. In the horizontal direction, we
split our sample into three velocity dispersion bins. Vertically, we
split the galaxies in each $\sigma$ bin by their activity type from
Starforming to quiescent (c.f. Figure \ref{fig:ur_csigma}). The contours are 25\% and
50\% probability contours respectively. Note that the area of these
contours can be misleading; only a small fraction of early-type
galaxies is in classes other than the quiescent bin. In order to
guide the eye, we indicate the fraction of galaxies in each mass bin
(low, intermediate and high) that are in each emission line class.
For example, in the high-mass regime, only a tiny fraction of
early-types are classified as Starforming (most likely
misclassifications), while the almost 90\% are quiescent;
nevertheless the contours occupy similar areas on the $t_{y}$-
$f_{y}$ plane. \label{fig:ty_fy_contours}}

\end{center}
\end{figure*}

What needs to be shown is twofold. First, solid evidence for the
presence of an evolutionary sequence for the low/intermediate-mass
objects is required. To this end, we have derived detailed star
formation histories using the method presented in section
\ref{sec:method} focusing on the epochs and mass fractions of the
most recent starburst. This allows us to investigate whether the
galaxies of the various sub-classes in our sample truly belong to the
same object type at different evolutionary stages. Second, the
interaction between star formation and nuclear activity along this
sequence must be investigated, with the aim to understand whether and how
the two processes are related. These are addressed in the following
sections.

\section{The evolutionary sequence}
\label{sec:results}

Following the discussion of the previous section we divide our sample
in three velocity dispersion bins ($\sigma\leq 100$,
$100<\sigma\leq 200~\mathrm{kms^{-1}}$, $\sigma>200~\mathrm{kms^{-1}}$) representing low-mass,
intermediate-mass, and massive early-type galaxies. Within each
$\sigma$ bin we maintain the 5 sub-classes: star forming, transition
object, Seyfert, LINER, and quiescent. Within each of the 15 resulting
sub-classes we stack the probability distributions for the star
formation histories obtained in the analysis of
Section~\ref{sec:method}. We marginalise over all parameters except
the mass-fraction $f_y$ and age $t_y$ of the starburst component. The
resulting contours of the probability distributions in the $t_y$-$f_y$
plane are shown in Fig.~\ref{fig:ty_fy_contours}. The panels are
analogues to panel (d) of Figs.~\ref{fig:young_example}
and~\ref{fig:old_example}. The five columns show the various emission
line classes, the rows are the three $\sigma$ bins. A bar in each
panel indicates the relative fraction of objects in this particular
sub-class. 

\subsection{Evidence for the time sequence}
\begin{figure*}
\begin{center}

\includegraphics[angle=90, width=0.49\textwidth]{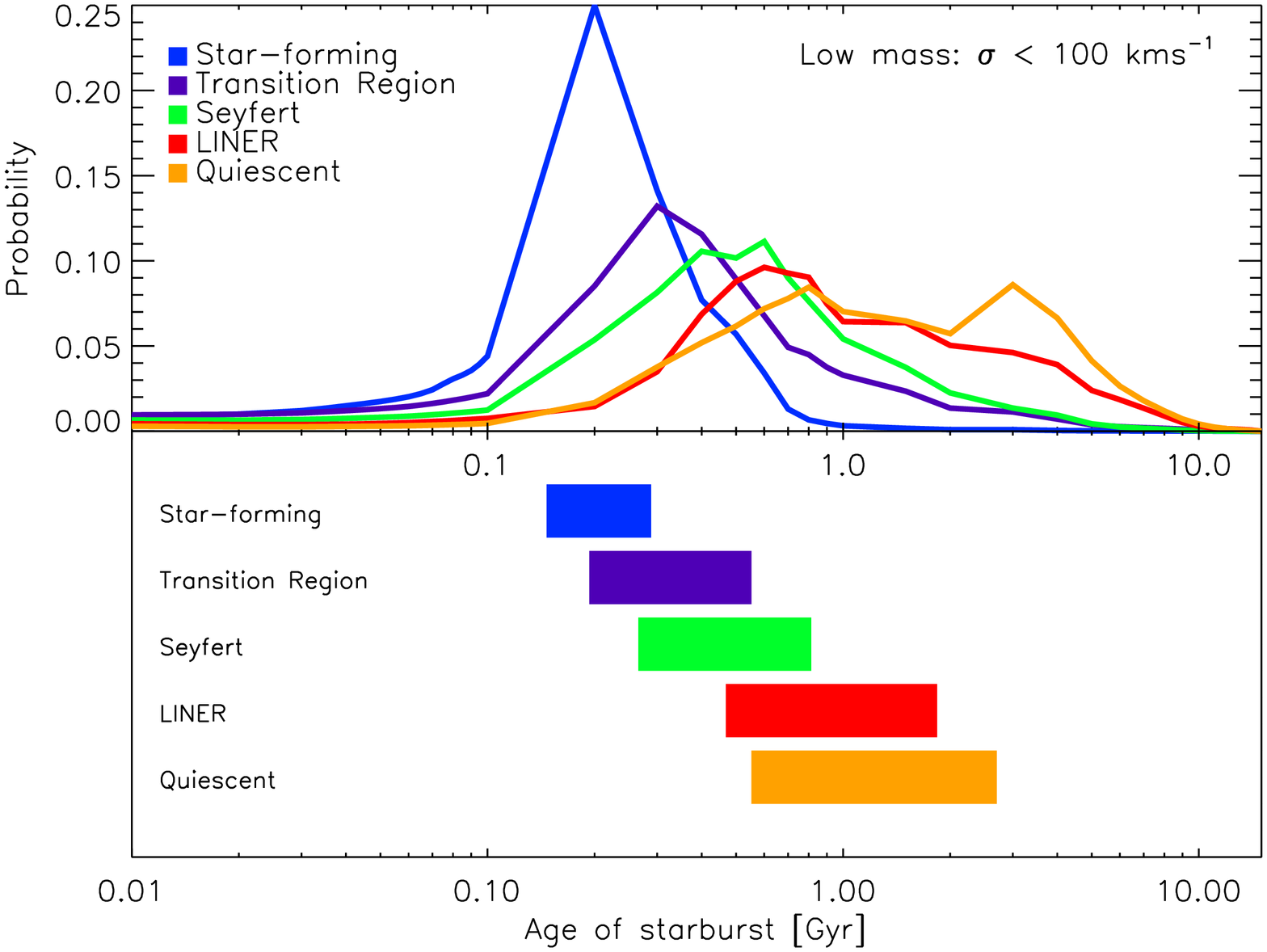}
\includegraphics[angle=90, width=0.49\textwidth]{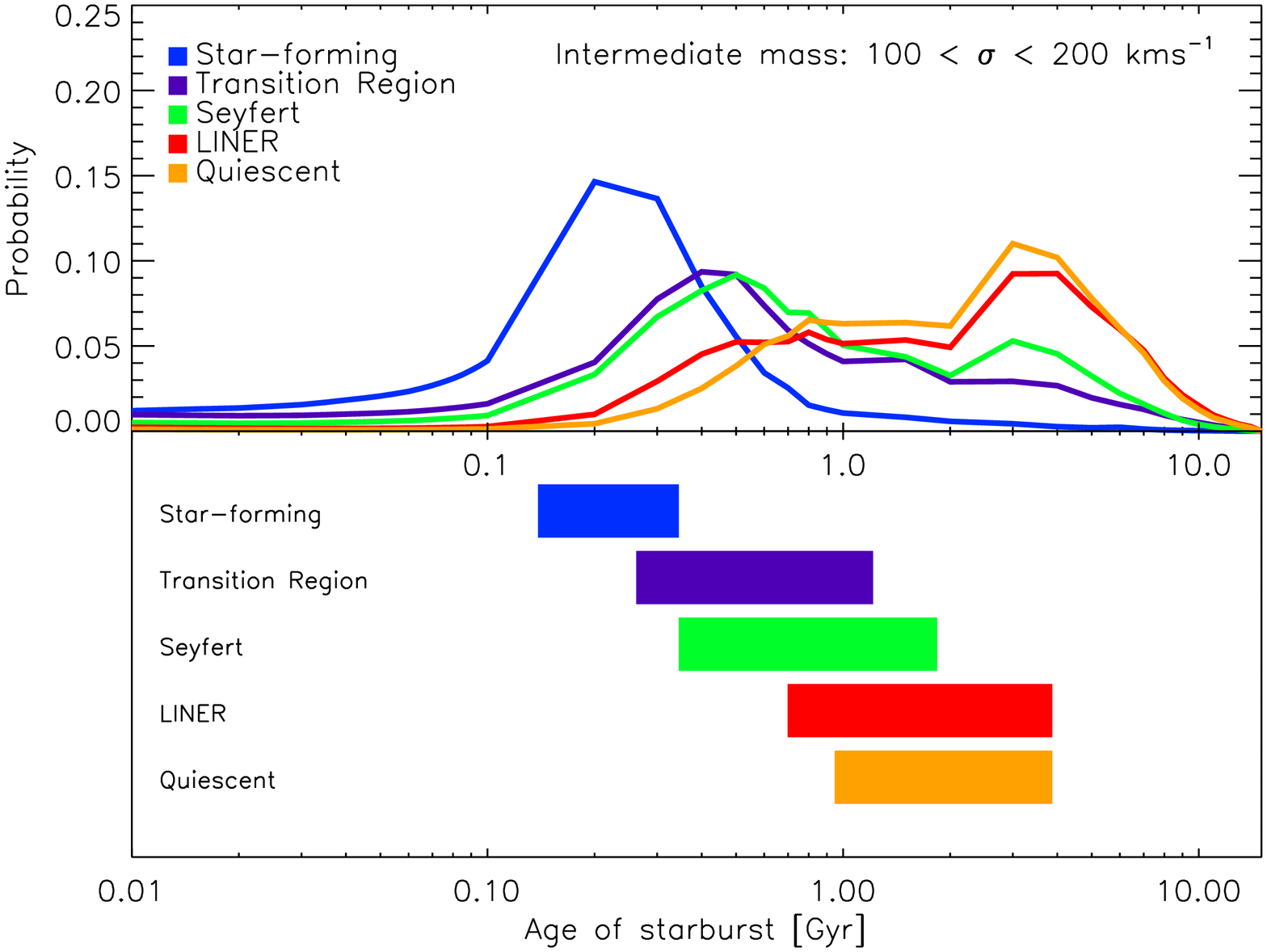}
\caption{The time sequence in the low- and intermediate mass bin:
  In the top of each panel, we show the
  stacked, normalised probability distribution functions for the
  time elapsed since the start of the starburst $t_{y}$. In the
  bottom, we indicate where on the the $t_{y}$ axis 50\% of the
  probability lie. We give the numbers for this in Table
  \ref{tab:time_sequence}.
  \label{fig:timeline}}

\end{center}
\end{figure*}

The results shown in Fig.~\ref{fig:ty_fy_contours} allow one immediate
conclusion. The star formation histories derived for the low-mass and
intermediate-mass objects in our sample galaxies reveal clear evidence
for the occurrence of a starburst between 100 Myr and 1 Gyr in the
past involving a typical mass fraction of 1 to 10 per cent. This reinforces
the interpretation of deviations from the red sequence in the
colour-$\sigma$ diagram (see Fig.~\ref{fig:ur_csigma}) to be caused by
recent rejuvenation events. 
Key are the epochs and mass fractions of this rejuvenation for the
various emission line classes. While the starburst ages clearly
increase from star forming via AGN dominated to quiescent in a
continuous sequence from $\sim 100$ Myr to beyond 1 Gyr, the mass
fractions remain the same. This points to the various emission line
classes having undergone the equivalent rejuvenation event in the
past, and the only feature differentiating the various classes is the
time elapsed since that star formation episode. Hence, the objects of
the various emission classes in the panels along one row in
Fig.~\ref{fig:ty_fy_contours} are identical in terms of mass,
morphology and star formation history, except that the epoch of the
last starburst varies. This means we have resolved an evolutionary
sequence, and what we are seeing is the process of rejuvenation in
early-type galaxies at different evolutionary stages. This time
sequence includes a phase of AGN activity along the way.

The fraction of objects taking part in this transformation is strongly
dependent on mass. In the lowest mass bin about one third of the
objects shows emission lines and is caught in the process of this
transformation. This fraction drops to 13 per cent in the intermediate
mass bin containing galaxies with $100\leq \sigma<200~\mathrm{kms^{-1}}$. In the
highest bin for objects with velocity dispersion above $200~\mathrm{kms^{-1}}$the
evolutionary sequence is not detected. At the high-mass end, the vast
majority of galaxies is quiescent, and no recent starburst within the
past few Gyrs has occurred. The small fraction of active galaxies is
by far dominated by LINER-like emission, with some transition region
objects that are in fact closer to the properties of LINERs (see
section \ref{sec:liner_mis}).

\subsection{Timescales along the sequence}

\begin{table*}
\caption{The Time Sequence: The Age of the Last Episode of Star
Formation as a Function of Mass and Emission Line Class}
\label{tab:time_sequence}
\begin{tabular}{@{}ll|lllll}
\hline
Mass Regime & Velocity Dispersion & Star Formation & Transition Region & Seyfert AGN & LINERs & Quiescent\\
& $\mathrm{kms^{-1}}$ & Myr & Myr & Myr & Myr & Myr \\
\hline
\hline
Low mass & 0-100 & 145-280 & 195 - 550 & 270-800 & 470 - 1800 & 550-2700\\
Intermediate mass & 100-200  & 140-350 & 260 - 1200 & 350-1800 & 700 - 4000 & 950-4000\\
\hline
\end{tabular}
\begin{flushleft}
The time intervals given in this Table correspond to the 50\%
probability intervals from Figure \ref{fig:timeline}.
\end{flushleft}
\end{table*}

For the low- and intermediate-mass objects ($\sigma <
200~\mathrm{kms^{-1}}$) we can now determine the relative and absolute
timing of this transformation process. We marginalise over the
mass-fraction and compute the stacked probability distributions of
starburst ages for the various emission line classes. The result is
shown in Figure \ref{fig:timeline} for the low- and intermediate-mass
early-type
galaxies in the left-hand and right-hand panels, while in Table
\ref{tab:time_sequence}, we give the figures in Myrs.
In both mass bins, we can see the shift in the age of the last episode
of star formation in a sequence as could already be seen from Figure
\ref{fig:ty_fy_contours}. The star forming objects are clearly the
ones most recently
rejuvenated with starburst ages around $150-300\;$Myr, with the latter
increasing along the sequence from transition region through Seyfert and
LINER to quiescent. 

A global interpretation yields the
following picture: The most common starburst age $t_y$ for the transition region objects is
around 300 to $500\;$Myr. Seyfert activity appears further 100 to
$300\;$Myr later with starburst ages peaking at about $600\;$Myr. This
phase is followed by LINER and quiescent with separations of a few
$100\;$Myr, respectively. The widths of the distributions increase
along this sequence from about $200\;$Myr for the star forming objects
to $400-600\;$Myr in the transition region and Seyfert objects. The
phases involving nuclear activity, transition and Seyfert, appear to
be of comparable length in time and must be a factor two to three
longer than the star forming one. Note that the widths of LINERs and
quiescent objects are much less well determined as they extend into
the regime of old ages beyond $1\;$Gyr, in which the age resolution is
significantly reduced.

The transition region and the Seyfert phases seem to overlap
considerably, which suggests that objects at this point along the
time sequence may switch back and forth between Transition
Region (a mixture of AGN and star formation) and Seyfert phase
(pure AGN). If this is the case, the Seyfert phase represents
episodes of enhanced accretion onto the black hole between phases with
lower levels of nuclear activity together with low residual star
formation. The star formation we see in the Transition Region objects
may even be \textit{induced} by the (potentially) recurrent Seyfert
state as the jets and outflows pile up gas before finally pushing it
out \citep{1998A&A...331L...1S, 2005MNRAS.364.1337S}.

Toward the end of the sequence, the transition from the
high-ionisation Seyfert phase to quiescence passes through a phase of
nuclear activity characterised by low-ionisation (LINER). In total,
the transformation process from blue cloud to red sequence passing
through nuclear activity takes about $800\;$Myr to $1\;$Gyr.

\subsection{Lifetimes and number fractions}
An important consistency check is the comparison of the lengths of the
various phases as derived from the stellar population ages with the
number of objects found in each phase. If the transition region and
Seyfert phases are equally long, the number of objects found in these
two classes must be the same. From Fig.~\ref{fig:ty_fy_contours} it
can be seen that this is not the case. We observe three to four times
more Transition Region objects than Seyferts. However, an unknown
fraction of the objects in the Seyfert phase must be Seyfert Type~I
rather than Type~II depending on the viewing angle from which the
nucleus is seen. Indeed, broad emission line objects like Type~I
Seyferts are not classified as 'galaxies' by the \textit{SDSS}
spectroscopic pipeline and hence are not included in our sample. Thus,
it is
plausible that the ostensible lack of Seyfert AGN in our sample stems
from this geometric effect, which puts object numbers and phase
lifetimes in very good agreement. It should also be noted that our
sample is not volume-limited.

\subsection{Dependence on mass}
The comparison between the lifetimes of the star forming and AGN
phases is more complex, as an additional mass dependence comes into
play. 
The fractions of the active early-type galaxy population in a given
mass bin that are classified as transition region and Seyfert objects decrease by
about a factor 2.5 from the low to the intermediate-mass bin, while
the number of star forming objects decreases by a factor 7. This
results in a factor 3.5 more transition region objects than star
forming ones at intermediate masses, while their fractions are
comparable at low masses. As the transition region phase lasts about
two to three times longer than the star forming one in both mass
regimes this suggests that not all low-mass objects go through the
time sequence. Hence star formation is not followed by nuclear
activity in all cases, and the true proportions between star forming
and transition region objects for the SF-AGN sequence may be observed
only at intermediate mass. This implies the presence of a tail of
young quiescent objects in the low-mass bin, which is indeed indicated
by the probability distributions in Fig.~\ref{fig:timeline}.

Still, the ratio transition region to star forming objects seems
slightly too high at intermediate masses. The reason for this may be
that some fraction of transition objects are misclassified and are
actually LINERs (see \ref{sec:liner_mis}) This would also explain why transition
objects and Seyfert appear less well separated at intermediate mass.

Within these uncertainties, the fraction of galaxies that experiences
residual star formation, and transitions from starburst to quiescence
via nuclear activity drops by about a factor two to four between the
low-mass and the intermediate-mass bins. The highest mass bin
($\sigma>200~\mathrm{kms^{-1}}$) concludes this trend, as no such
sequence is detected for these masses.
This is very well supported by the fact that the apparent dichotomy in the starburst age distribution within the quiescent early-types galaxies (left-hand panel of Fig.~\ref{fig:timeline}) is considerably weaker at intermediate masses (right-hand panel), and disappears entirely at the high-mass end (see Fig.~\ref{fig:ty_fy_contours}).

Finally, it is interesting to note that, while the vast majority of
LINERs detected in the low-mass bin appears to take part in the
evolutionary sequence, the presence of a second independent LINER
population seems to emerge at intermediate
masses. Fig.~\ref{fig:timeline} reveals a clear dichotomy with a fair
fraction of LINER-like objects that does not show signs of recent
rejuvenation. This trend becomes even more evident at the highest
masses, where the largest fraction of LINERs is detected (8 per cent),
which contain only old stellar populations (see
Fig.~\ref{fig:ty_fy_contours}).

\subsection{Dust properties along the sequence}

\begin{figure}
\begin{center}

\includegraphics[angle=90, width=0.5\textwidth]{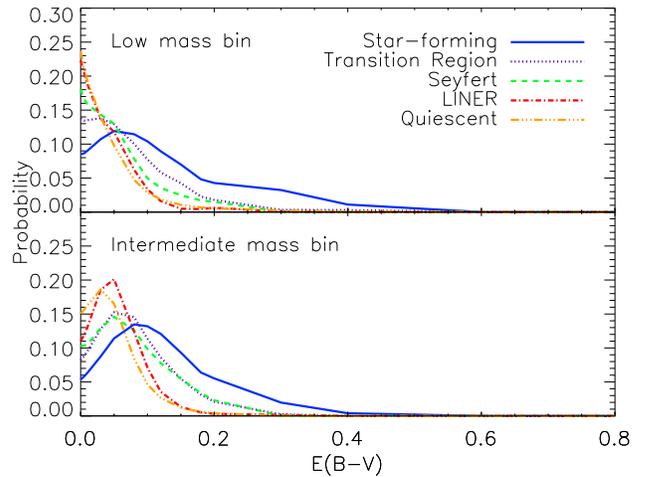}
\caption{This plot shows the stacked, normalised probability
  distribution functions for the dust parameter $E(B-V)$ for the
  low and medium mass bin. Starforming early-types are the dustiest with the
  Transition Region and Seyfert early-types being less dusty and the
  LINERs and quiescent early-types containing even less dust.
  \label{fig:dust_plots}}

\end{center}
\end{figure}

Fig.~\ref{fig:dust_plots} shows the stacked, normalised probability
distribution functions for the dust parameter $E(B-V)$ obtained
through our SED fitting algorithm for the low- and intermediate-mass
bins. The different line-styles are the various emission line classes
(see labels). As ought to be expected, dust content decreases along
the sequence discussed above with highest dust attenuation in the star
forming objects ($E(B-V)\sim 0.1$) and the lowest in the quiescent
ones ($E(B-V)\la 0.05$). This internal consistency further strengthens
the conclusion that we are indeed observing a time sequence and are
witnessing the transformation of early-type galaxies from star
formation via AGN to quiescence. In order to mimic a time sequence in
the observed $u-r$ colours with a variation in the amount of dust rather
than in age, the trend in $E(B-V)$ would have to be the opposite of what we observe.

\section{Discussion}
\label{sec:discussion}
With Figure \ref{fig:ur_csigma} in Section~\ref{sec:redblue} we show
that early-type galaxies of different emission line classes occupy
different loci in the colour-$\sigma$ diagram. Star forming
early-types are the bluest deviating most from the red sequence. Once
some AGN contribution appears in the optical emission lines, the
colours dramatically redden. In Section \ref{sec:results}, we
demonstrate quantitatively that there is indeed strong evidence for
the existence of an evolutionary sequence of early-type galaxies
moving from the blue cloud to the red sequence. We show that on this
path the galaxies go through a number of phases starting at star
formation followed by a transition region of star formation and AGN, a
Seyfert AGN phase and a LINER phase before finally settling to total
quiescence and passive evolution on the red sequence. This process is
currently building the low-mass end of the red sequence in the low
redshift universe. In this section we discuss the astrophysical
implications of this finding, the role of LINER-like early-type
galaxies, possible trigger mechanisms of this sequence, and follow-up
observational campaigns in the context of the time sequence we have
derived.

\subsection{The starburst-AGN sequence}
A clear implication of our results is the existence of a link between residual starburst and nuclear activity at recent epochs in low- and intermediate-mass early-type galaxies. Whatever triggers the starburst, it leads also to an AGN, and the star forming phase reaches quiescence via an AGN phase. We see the starburst fading while the AGN rises. This leads to two possible scenarios depending on whether the two processes, star formation and AGN, are connected.

\subsubsection{AGN feedback or black hole growth in action?} 
If the fading of star formation and nuclear activity are linked, then we are observing AGN feedback in action. The appearance of the
AGN in the transition region and Seyfert phases are concurrent with
the end of star formation, so it is plausible that these two phenomena
are connected such that the rise of the AGN suppresses and terminates
star formation by heating and expulsion of the remaining gas.

Alternatively, it is possible that starburst and AGN actually do not
interact with each other, even though they may be triggered
by the same global event. In this case star formation fades simply owing to supernova feedback and/or the exhaustion of cold gas. 
Most interesting in the framework of this scenario is our finding that the AGN rises while star formation declines. This shows that in this case AGN activity must be delayed substantially with respect to the starburst by several hundred Myr and may actually be triggered by star formation.
This implies that the early-type galaxy or its progenitor hosting the
starburst is just in the process of building up its central super-massive black hole.

\subsubsection{Gas depletion timescale: in favour of AGN feedback}
In order to distinguish these two scenarios, we compare the observed timescale for the fading of star formation with simple theoretical estimates based on the Schmidt-Kennicutt law \citep{1959ApJ...129..243S, 1998ApJ...498..541K}. The latter relates the star formation rate with the ratio of gas mass surface density and the dynamical timescale of the system scaled by an efficiency $\epsilon$. We assume that the dynamical timescale of star formation is on typical galaxy timescales of $\sim 0.1-0.5$ Gyr, as observed in gas discs in elliptical galaxies \citep{2002AJ....124..788Y}. 
We can exclude shorter dynamical timescales caused by a concentration of the star burst activity to small radii, as the mean petrosian half-light radii of the \textit{u}-band light in the starforming
early-types ($2.25\arcsec$) imply that star formation in these
objects is extended over several kpc.
Then to deplete any gas reservoir would take on the order of many Gyrs. 

This timescale is significantly longer than the fading timescale deduced here for our sample galaxies. In our modelling analysis we find that the recent rejuvenation event is best modelled by a exponentially declining star burst with e-folding timescale around $100\;$Myr. Thus, the gas reservoir cannot get depleted \emph{solely} via self-regulated star formation. An additional heating source is required. The fact that we find an AGN phase subsequent to the starburst strongly suggests that the AGN is (at least partly) responsible for the suppression of star formation by depleting the gas reservoir accelerating the transition from the blue cloud to the red sequence. A more detailed investigation of this process is beyond the scope of this paper. It will be explored in more detail in a forthcoming work (Kaviraj et al., in prep).

This argument strongly supports the AGN feedback interpretation of the time sequence. This result has important consequences for our understanding of galaxy formation. It is empirical proof that AGN feedback, the existence of which is now postulated in most recent models of hierarchical galaxy formation (see Introduction and references therein), occurs in early-type galaxies at late stages of their evolution. 
Beyond this important conclusion, our sample that is resolving the actual sequence of AGN feedback provides further insight into this process. We find its overall timescale to be of the order of a Gyr. Moreover, we observe a substantial delay between fading of star formation and rise of the AGN by a few hundred Myr, which might reflect the timescale required for the black hole to build up the accretion disc. This implies that AGN feedback is not well approximated with an immediate truncation as implicitly assumed in current hydrodynamical simulations \citep{2005MNRAS.361..776S}.

\subsection{Obscured AGN?}
A key result is that we see the AGN rising while star formation fades.
As absolute O[III] luminosities increase along the sequence, it can be excluded that the AGN is simply out-shined by the starburst during the star forming phase. However, it is still possible that the AGN is not observed during the starburst, because it is highly obscured in the beginning. In this case, star formation and AGN would coexist from the beginning, and while star formation fades, the AGN remains active. This does not invalidate but rather strengthen the interpretation of AGN feedback occurring in our sample galaxies.

\subsection{Progenitors and trigger mechanisms}
The youngest early-type galaxies on the evolutionary sequence we
resolve are the star forming ones with starburst ages around
$100-200\;$Myr. The progenitors of these objects within the sequence
must have stellar populations younger than $200\;$Myr. The fact that
we do not observe these objects implies that they are excluded through
our selection criteria, hence they most likely do not have early-type
morphology. 

\subsubsection{Prodigious star formation in the recent past}
As we model the starburst as composite stellar population with exponentially declining star formation histories, these progenitors must have had significantly higher star formation rates in the past. 
The typical star forming early-type galaxy has a mass of about
$5 \times 10^{10}  M_{\odot}$ and has formed about 5\% of its stellar mass in
the current burst (i.e. $2.5 \times 10^9  \rm M_{\odot}$) over the last
$100\;$Myr. This corresponds to an \textit{average} star formation
rate of $25~M_{\odot} yr^{-1}$, resulting in an infrared
luminosity $ L_{\rm IR}$ of $1.25 \times 10^{11}  L_{\odot}$, well
above the threshold for so-called luminous infrared galaxies
(LIRGs). The star formation rate was undoubtedly higher in the past as
the best fit decay timescale $\tau$ is around 100 Myr. Such an
exponentially decaying star formation rate can easily boost the past
$L_{\rm IR}$ into the category of ultra-luminous infrared galaxies
(ULIRGs). 

We find that the most luminous starforming early-type
galaxy detected by the \textit{Spitzer Space Telescope} in the
\textit{SDSS}-SWIRE overlap has $\mathrm{L_{FIR}} = 5.7-7.4 \times
10^{10}~\mathrm{L{\odot}}$ \citep{1996ARA&A..34..749S} implying a star formation rate of $13~ M_{\odot} yr^{-1}$ based on the
relationship of \cite{2004ApJ...606..271G}. This galaxy is thus
just slightly below the cut defining a LIRG of $L_{\rm IR}  =
10^{11} L_{\odot}$ at the time it is observed.

We conclude that the progenitors of the objects we observe on the
SF-AGN sequence were most likely ULIRGs. This fits well with the evidence that the
local ULIRG population most likely will evolve into intermediate-mass
early-type galaxies \citep{2001ApJ...563..527G}. Dasyra et al.~(2006a, 2006b)\nocite{2006ApJ...638..745D,2006ApJ...651..835D} and \citet{2006ApJ...643..707V} have analysed HST H-band imaging and VLT spectra of low redshift ULRIGs and QSOs and
concluded that while most of them are ongoing mergers, their
kinematic, structural, and photometric properties
are consistent with them becoming elliptical galaxies and settling
on the fundamental plane. Most relevant here is that they
find their stellar velocity dispersions to have typical values of
around $150~\mathrm{kms^{-1}}$ in overlap with the typical
velocity dispersions of our active galaxies that mark the
transformation process from blue cloud to red sequence. 

The appearance of an AGN phase several hundred Myr \textit{after} the U/LIRG phase may explain why most local ULIRGs do not appear to contain a luminous AGN \citep{1998ApJ...498..579G}.

\subsubsection{Merger as main trigger}
The origin of ULIRG activity is thought to be a galaxy merger \citep[e.g.][]{2001ApJ...563..527G}. This implies that we are possibly seeing a slightly more advanced stage of a gas-rich merger where the morphology has already settled into a spheroid, and the infrared luminosity has already decayed below the LIRG threshold.

Because of our sample selection, all objects discussed here are morphologically early-type galaxies without spiral or disc structure as far as detectable in the \textit{SDSS} images. Elliptically shaped galaxies with minor morphological perturbations such as tidal features or dust lanes are {\em not} excluded. This selection should in principle enable us to investigate the role of morphological disturbance, hence merger and galaxy interaction activity, along the sequence. However, the fraction of disturbed early-type galaxies in our sample is quite small. We have not found any compelling trend between morphological irregularities and star formation/nuclear activity, most likely because SDSS imaging does not provide the necessary sensitivity in terms of depth and spatial resolution. Deeper imaging, preferably at higher spatial resolution, is clearly needed to identify signatures of recent merger activity in our sample objects (see e.g., \citealt{2005AJ....130.2647V} and \citealt{2006ApJ...640..241B}).

\subsection{Eddington ratios along the sequence}

\begin{figure*}
\begin{center}

\includegraphics[angle=90, width=0.49\textwidth]{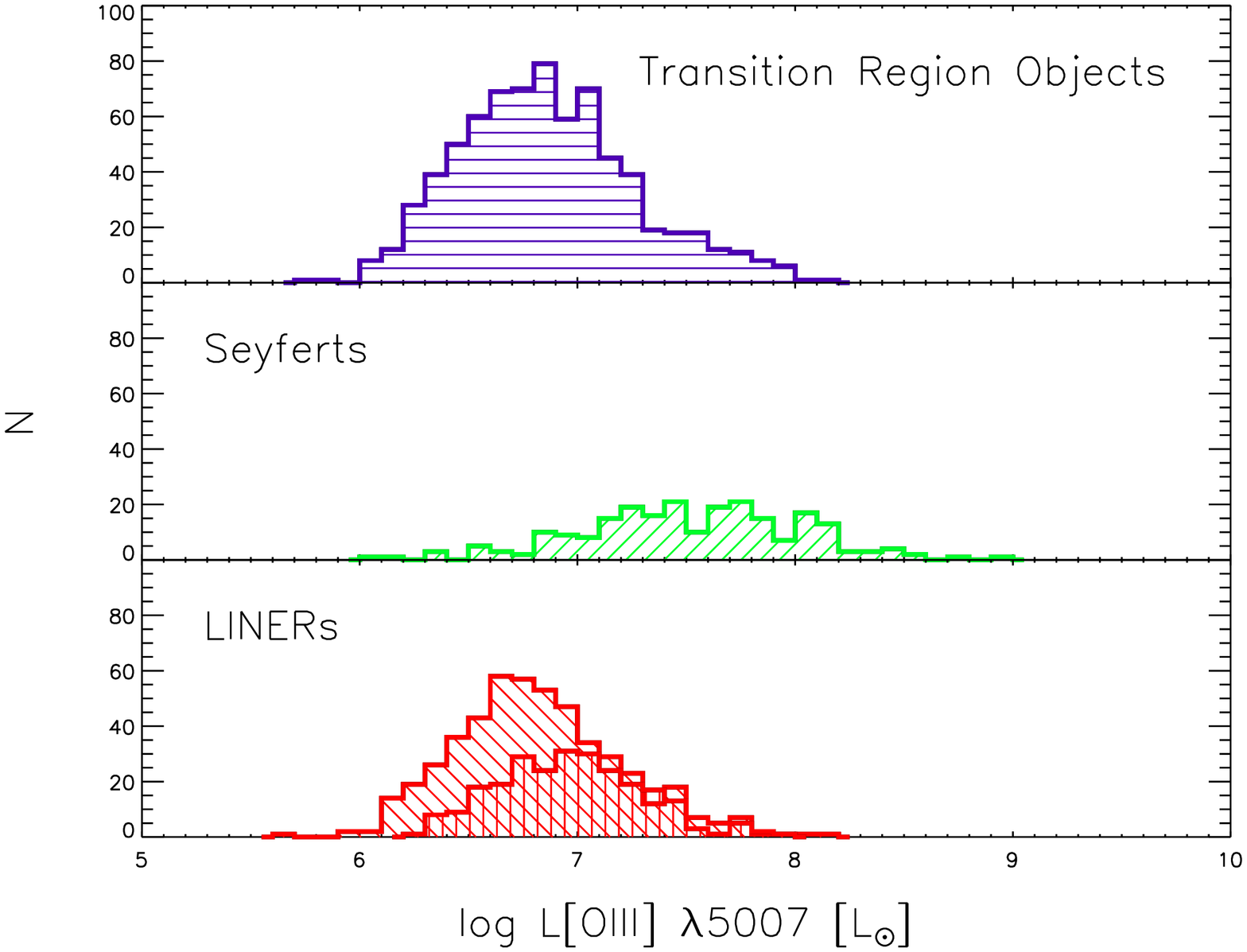}
\includegraphics[angle=90, width=0.49\textwidth]{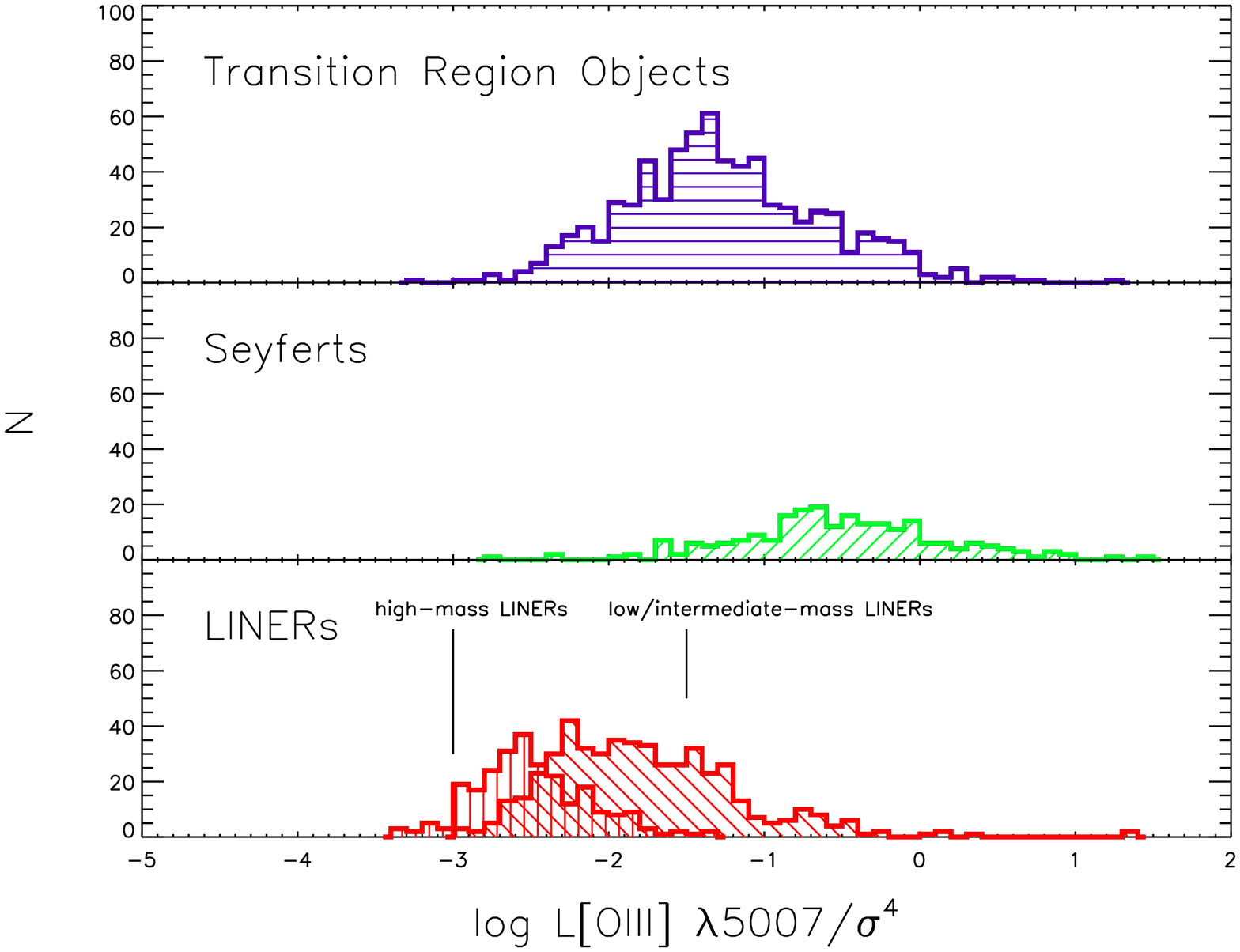}
\caption{In the left-hand part of this Figure, we show the histograms
  for the L[OIII] $\lambda5007$ luminosities of our AGN and transition
  region objects. [OIII] is a tracer of the accretion rate
  (\citealt{2004ApJ...613..109H}) and so is an upper limit on the
  accretion rate of the transition region objects. In the left-hand
  part, we show
  the histogram of the quantity L[OIII]/$\sigma^4$, which is
  proportional to the Eddington ratio
  (\citealt{2006MNRAS.372..961K}). In each part, we go along the
  sequence from early times to late. First, we show the transition
  region objects, then the Seyferts and then the LINERs. We split the
  LINERs into two; while the
  low- and intermediate-mass LINERs and the high-mass LINERs have
  similar [OIII] luminosities, their Eddington ratios are very
  different. The low/intermediate-mass LINERs form the low luminosity
  and high Eddington ratio end of the Seyfert distribution, as would be
  expected from their place in the time sequence. The high-mass LINERs
  have Eddington ratios an order of magnitude below their lower-mass
  counterparts. Along the sequence, both absolute L[OIII] and
  Eddington ratio rise, peak and then decline over the time domain
  covered by the time sequence. \label{fig:agn_luminosity}}

\end{center}
\end{figure*}

In the following we discuss the nature of accretion onto the supermassive black hole along the AGN feedback sequence.
In Figure \ref{fig:agn_luminosity}, we show the distributions of [OIII]
luminosity and the quantity L[OIII]/$\sigma^4$ for Seyfert, low-mass and high-mass LINERs. L[OIII]/$\sigma^4$ has been
introduced by \cite{2006MNRAS.372..961K} as a tracer of the
Eddington ratio assuming that [OIII] luminosity scales
with the AGN bolometric luminosity \citep{2004ApJ...613..109H},
and that $\sigma^4$ provides an estimate of the central super-massive black hole mass \citep{2000ApJ...539L..13G, 2000ApJ...539L...9F}.
Based on this quantity \cite{2006MNRAS.372..961K} show that the fundamental difference between Seyferts and LINERs in the \textit{SDSS} data base is the accretion efficiency, with the LINERs being at lower Eddington ratios than the Seyferts. Since the transition region objects [OIII] line luminosities have some
contribution from star formation, we can treat them as upper limits on both the accretion rate and efficiency.

We find that the Seyfert early-types in our sample are high-luminosity
AGN with high L[OIII] luminosities and high L[OIII]/$\sigma^4$ ratios,
indicating that they are strongly accreting systems relatively close
to the Eddington limit. The LINERs, instead, have lower L[OIII]
luminosities and lower Eddington ratios. There is an interesting trend
with mass, however. The massive LINERs exhibit the lowest accretion
rates, while the low- and intermediate-mass objects are found between
massive LINERs and Seyferts. Hence our low-mass and high-mass
LINER early-type galaxies are objects in different states of
nuclear activity. The former represent probably the end stage of the
SF-AGN sequence. If the Seyfert phase is a high-accretion phase during
which the AGN feedback process occurs, then the transition to the
LINER phase is the point at which the supply of gas has been largely
used up by star formation and/or accretion onto the black hole. The
massive LINERs characterised by low Eddington ratios, instead, might
be responsible (or at least part of) the rigourous AGN feedback
process in massive early-type galaxies that avoid rejuvenation in the
first place.
Figure \ref{fig:agn_luminosity} illustrates the extended rise and decline in both accretion rate
and Eddington ratio over the time domain covered by the time sequence. While the transition region to Seyfert 
evolution brings with it an increase in accretion, so the progression to a LINER is accompanied by a decline. 

\subsection{The role of LINERs}
\label{sec:liners}
Fig.~\ref{fig:ty_fy_contours} shows that the fraction of LINERs
observed increases with increasing galaxy mass, while the fraction of
galaxies undergoing the SF-AGN sequence decreases. Hence, LINERs
appear to come in two flavours. Some are part of the evolutionary
sequence of low- and intermediate-mass galaxies, but the majority is
found in the massive, quiescent early-type galaxies of our sample. 

It should be emphasised that the LINER-like emission observed here is
most likely caused by nuclear activity. The extended LINER-like emission not
associated with an AGN found by \citet{2006MNRAS.366.1151S} in very
nearby early-type galaxies is significantly weaker (typically with total
L[OIII] $ < 10^{6} \rm L_{\odot}$; c.f. Figure
\ref{fig:agn_luminosity}) and would remain undetected in our sample.

\subsubsection{LINERs in low and intermediate masses}
As discussed in the previous sections, the early-type galaxies with
LINER emission appear on the evolutionary sequence of low- and
intermediate-mass objects between Seyfert-like nuclear activity and
quiescence. 
This means that early-type galaxies transition from a high-accretion
regime characterised by Seyfert-emission to a low accretion regime
characterised by LINER-emission \citep{2006MNRAS.372..961K} as they
proceed to fade onto the red sequence. This can be interpreted as the
AGN running out of fuel. This is well consistent with the study of
\citet{2006ApJ...648..281Y}, who find that many post-starburst
galaxies show strong [OII] emission, but no H$\alpha$,
indicating no further ongoing star formation, which they associate
with LINERs.

\subsubsection{LINERs in massive galaxies}
Already in the intermediate mass bin, most LINERs appear to be associated
with stellar quiescence (Figure \ref{fig:ty_fy_contours}), and none of the
LINERs in the highest mass bin are connected with rejuvenation. This
might be AGN feedback in its most rigourous form and is consistent
with the fact that the efficiency of AGN feedback must increase with
increasing galaxy mass in the models in order to reproduce observed UV
colours \citep{2006Natur.442..888S}. While late 
star formation seems to be terminated by nuclear activity in low-mass
early-type galaxies, the AGN in massive galaxies may prevent star
formation in the first place. These
massive galaxies most likely have experienced strong AGN feedback at
high accretion nuclear activity at earlier epochs. Indeed,
observational evidence for the existence of this process has recently
been found
\citep{2006astro.ph.11724K,2007MNRAS.376..416R,2007arXiv0705.2832D}.

This LINER-like activity may play the crucial role in
maintaining an atmosphere of hot X-ray emitting gas which can shield
the galaxy from gas accretion and cooling and so prevent any further
star formation as suggested by \citet{2004MNRAS.347.1093B}. Such hot
gas may then trigger \textit{extended} LINER-like emission via thermal conduction of
self-irradiation as has been seen in massive nearby elliptical
galaxies (see e.g \citealt{2007NewAR..51...18S} and references
therein), although it is likely that the
majority of the LINER emission in our sample is associated with (more
luminous) nuclear activity. 

\subsubsection{Mis-classified LINERs}
\label{sec:liner_mis}
We note that some transition region early-types are found in the
highest mass bin, which show no
evidence for a star formation episode (see
Fig.~\ref{fig:ty_fy_contours}). These transition region objects, being
the result of a combination of star formation and AGN activity, have
no corresponding starforming or Seyferts at the same mass and do not
seem to take part in the AGN-SF time sequence. Instead, their
properties are far more similar to those of high-mass LINERs.
The
most likely explanation for this apparent contradiction is that the
extreme starburst line of \cite{2001ApJ...556..121K} led to the
mis-classification of some LINERs as transition region objects. An
improved classification scheme including the low-ionisation lines may
help here but is beyond the scope of this paper.

\subsection{Follow-up observations}
The sample presented in this paper is the ideal laboratory for more
detailed follow-up studies of the link between star formation and
nuclear activity in early-type galaxies. Multi-wavelength data will
tighten constraints on the formation histories, the star formation
activities, dust contents, central black hole activities, and
accretion rates of our sample galaxies. This will provide deeper
insight into the physics involved.

By observing CO molecular transitions the fate of the molecular gas
can be explored. Such observations clarify whether it is really is the
feedback from AGN that quenches star formation, or whether this
decline in star formation is caused by the fuel supply being exhausted
by star formation itself. High-resolution radio continuum observations
at different frequencies provide important clues on the relative
contributions of star formation and the AGN along the sequence. The
relative importance and interaction between AGN and star-formation in
the galaxies along our timeline can be further addressed by the
analysis of the full SED in the mid-infrared wavelength
regime. Mid-infrared imaging helps to pin down the relative amounts of
cold, warm and hot dust, while spectroscopy can disentangle
contributions from PAH lines to any hotter dust components due to an
AGN. Finally, deep optical imaging would be very valuable and help to
determine the role of mergers and interactions along the AGN feedback
sequence.

Observations at the observatory of the Institut de Radioastronomie
Milimetrique (IRAM) and at the Very Large Array (VLA) to address part
of these questions have been taken in June and July 2007.

\section{Conclusions}
\label{sec:summary}
We study the relationship between star formation and nuclear activity
in early-type galaxies at recent epochs. The principle aim is to test
whether AGN feedback occurs in these objects, as invoked in most
recent models of hierarchical galaxy formation. 

For this purpose, we analyse a magnitude-limited sample of 48,023
galaxies drawn from the \textit{SDSS} DR4 in the redshift range of
0.05 to 0.1 limited to \textit{r} $<$ 16.8. We visually inspected all
galaxies in this sample identifying 15,729 morphological early-type
galaxies. This paper is the first of a project called MOSES:
\textbf{MO}rphologically \textbf{S}elected \textbf{E}llipticals in
\textbf{S}DSS. A detailed description of the catalogue can be found in
a companion paper (Schawinski et al., in preparation). Most
importantly, the MOSES sample is not biased against star formation and
nuclear activity thanks to the purely morphological selection
criterion. The SDSS photometry is matched to a number of other
surveys, namely \textit{GALEX}, \textit{2MASS}, and \textit{SWIRE},
yielding a wavelength coverage from the far-UV to mid-IR. Re-analysing
the \textit{SDSS} spectra, we measure emission line fluxes, stellar
absorption line indices and velocity dispersions. The contributions of
the stellar continuum and of the ionised-gas emission to the observed
galaxy spectrum are separated by fitting \textit{simultaneously}
stellar population templates.

18.4 per cent of the early-type galaxies in our sample show emission
lines, and we employ emission line ratio (BPT) diagrams to classify them into
starforming, AGN-SF composites, Seyfert AGN and LINERs. The fraction
of emission line galaxies increases with decreasing galaxy mass to 40
per cent at the low-mass end ($\sigma < 100~\rm{kms^{-1}}$). The
emission line classes are roughly evenly distributed between star
formation and AGN at intermediate ($100< \sigma\leq 200\;$km/s) and
low ($\sigma\leq 100\;$km/s) masses. At high masses
($\sigma>200\;$km/s), instead, only LINER-like emission is
detected. The H$\alpha$ star formation rates of those objects
classified as currently star forming range from
$0.1-25\;\rm{M_{\odot}yr^{-1}}$.

We find that the objects with emission are offset from the red
sequence and form a well-defined pattern in the colour-$\sigma$
diagram. Star forming early-types inhabit the blue cloud, while
early-types with AGN (Seyfert and LINER) are located considerably
closer to and almost on the red sequence. Star formation-AGN
composites are found right between these two extremes. There is a
sequence between the blue cloud and the red sequence from star forming
via transition region and Seyfert AGN and LINER to quiescence.

To analyse this transition in more detail, we have developed a method
for deriving star formation histories with particular focus on
quantifying the epoch and mass fraction of the last significant
episode of star formation. Novel about our approach is to use a
combination of  UV-optical-NIR photometry and stellar absorption
indices as observational constraints.

We find that low- and intermediate-mass early-type galaxies with
emission lines have experienced an episode of star formation less than
$\sim 1$ Gyr ago involving $1-10\;$ per cent of their total stellar
mass. This recent star burst is best modelled by a steeply declining
exponential with e-folding timescale around $100\;$Myr. Most
importantly, this timescale and the mass fraction involved is the same
for all the emission line early-type galaxies whether they are
starforming, Transition Region, Seyferts or LINERs. They are offset in
time, however, with the starforming being the youngest, and the
Seyfert/LINERs being the oldest. Quiescent galaxies (objects without
emission lines) have star burst ages larger than $1\;$Gyr, which sets
the timescale for the transition process we observe.
The duration of the various phases along this sequence agree
reasonably well with the number fractions observed, once geometric
effects of Seyfert activity are taken into account. Dust content
decreases along the sequence with highest dust attenuation in the star
forming objects ($E(B-V)\sim 0.1$) and the lowest in the quiescent
ones ($E(B-V)\la 0.05$). 

These results lead to the conclusion that we are observing an
evolutionary sequence and are witnessing the transformation of
early-type galaxies from star formation via AGN to quiescence. Along
this sequence, nuclear activity rises while star formation fades. We
discuss that gas exhaustion or self-regulation of star formation alone
does not suffice to explain this fading, which leads to the conclusion
that we have detected a sequence in which nuclear activity suppresses
star formation. We are seeing AGN feedback in action, while low- and
intermediate-mass early-type galaxies transition from the blue cloud
through the green valley, characterised by nuclear activity, onto the
red sequence.
The high-mass galaxies in our sample, instead, did not go through this
transition in the recent past. These objects have only LINER-like
emission and are not associated with rejuvenation. This indicates that
the AGN in massive galaxies may prevent star formation in the first
place, which would be AGN feedback in its most rigourous form. 

To conclude, we have found empirical evidence for the occurrence of
AGN feedback in early-type galaxies at recent epochs. The galaxy
sample presented here is the ideal laboratory for testing more
detailed physics of this process. A first important hint from this
work is that there is a substantial delay between fading of star
formation and rise of the AGN by a few hundred Myr. This implies that
AGN feedback is not well approximated with an immediate truncation as
implicitly assumed in current models.

\section*{Acknowledgements}
We would like to thank the referee for the thorough and prompt report.
We thank Andi Burkert, Stephen Justham, Sadegh Khochfar,
Chris Lintott, Lance Miller, Dimitra Rigopouplou, Clive Tadhunter and Roberto
Trotta for many stimulating discussions. We would also like to
thank Stephen Justham and Philipp Podsiadlowski for letting us
use their computing resources.

This project has been partly supported by grant BMBF-LPD 9901/8-111 of
the Deutsche Akademie der Naturforscher Leopoldina. SK acknowledges a
Leverhulme Early-Career Fellowship, a BIPAC fellowship and a Research
Fellowship from Worcester College, Oxford. SJJ was supported by
the Korea Research Foundation Grant funded by the Korean Government
(MOEHRD) (KRF- 2005-213-C00017). CM ackwnoledges the grant
MEIF-CT-2005-011566 of the Training and Mobility of Researchers
programme financed by the European Community.

The SDSS is managed by the Astrophysical Research Consortium (ARC) for
the Participating Institutions. The Participating Institutions are the
University of Chicago, Fermilab, the Institute for Advanced Study, the
Japan Participation Group, The Johns Hopkins University, the Korean
Scientist Group, Los Alamos National Laboratory, the
Max-Planck-Institute for Astronomy (MPIA), the Max-Planck-Institute
for Astrophysics (MPA), New Mexico State University, University of
Pittsburgh, University of Portsmouth, Princeton University, the United
States Naval Observatory, and the University of Washington.

This publication makes use of data products from the Two Micron All
Sky Survey, which is a joint project of the University of
Massachusetts and the Infrared Processing and Analysis
Center/California Institute of Technology, funded by the National
Aeronautics and Space Administration and the National Science
Foundation.

\bibliographystyle{mn}
\bibliography{bibliography}

\bsp

\label{lastpage}

\end{document}